\documentclass[namedreferences]{SolarPhysics}
\usepackage[optionalrh,solaenum]{spr-sola-addons} 
\usepackage{graphicx}                    
\usepackage{color}                       
\usepackage{url}                         

\begin{document}

\begin{article}

\begin{opening}

\title{Evolution of active and polar photospheric magnetic fields during the rise of Cycle 24 compared to previous cycles}

%
\author{G.J.D.~\surname{Petrie}      
       }

%

%
  \institute{National Solar Observatory, Tucson, AZ 85719, USA
                     email: \url{gpetrie@noao.edu}\\
             }

\begin{abstract}
The evolution of the photospheric magnetic field during the declining phase and minimum of Cycle 23 and the recent rise of Cycle 24 are compared with the behavior during previous cycles. We used longitudinal full-disk magnetograms from the NSO's three magnetographs at Kitt Peak, the Synoptic Optical Long-term Investigations of the Sun (SOLIS) Vector Spectro-Magnetograph (VSM), the Spectromagnetograph and the 512-Channel Magnetograph instruments, and longitudinal full-disk magnetograms from the Mt. Wilson 150-foot tower. We analyzed 37 years of observations from these two observatories that have been observing daily, weather permitting, since 1974, offering an opportunity to study the evolving relationship between the active region and polar fields in some detail over several solar cycles. It is found that the annual averages of a proxy for the active region poloidal magnetic field strength, the magnetic field strength of the high-latitude poleward streams, and the time derivative of the polar field strength are all well correlated in each hemisphere. These results are based on statistically significant cyclical patterns in the active region fields and are consistent with the Babcock-Leighton phenomenological model for the solar activity cycle. There was more hemispheric asymmetry in the activity level, as measured by total and maximum active region flux, during late Cycle 23 (after around 2004), when the southern hemisphere was more active, and Cycle 24 up to the present, when the northern hemisphere has been more active, than at any other time since 1974. The active region net proxy poloidal fields effectively disappeared in both hemispheres around 2004, and the polar fields did not become significantly stronger after this time. We see evidence that the process of Cycle 24 field reversal has begun at both poles.
\end{abstract}

%

\end{opening}

%
\section{Introduction}
\label{s:introduction} 

Cycle 24 is different from the three previous cycles observed since the mid-70s in at least three important respects: (1) the weakness of the polar fields at the beginning of the cycle, (2) the low level of activity and the very slow increase in activity during the early years of the cycle, and (3) the strong asymmetry of the flux between the north and south hemispheres. The polar field is about 40\% weaker than the two previous cycles (Wang et al.~2009, Gopalswamy et al.~2012) and polar coronal hole areas during the recent Cycle 23 minimum were about 20\% smaller (Wang et al.~2009). Also the heliospheric flux was at its weakest level since measurements began in 1967 (Sheeley~2010). At the same time low-latitude coronal holes were larger and more numerous than during the previous minimum (e.g., Lee et al.~2009). These phenomena are likely to be related to the weakness of the polar fields, which play a critical role in determining the structure of the coronal and heliospheric field, particularly at times of low activity. The weak polar fields have been linked by Wang et al.~(2009) to a 20\% shrinkage in polar coronal-hole areas and and a reduction in the solar-wind mass flux over the poles. Lee et al.~(2009) showed that the weak polar fields are likely to be responsible for the low interplanetary mean field (IMF) and a nonuniform distribution of this field at 1 AU at low to mid latitudes during Cycle 23. It is well known that the coronal streamer structure and the heliospheric current sheet only became axisymmetric in the equatorial plane after sunspot numbers fell to unusually low values in mid-2008 (Wang et al. 2009, Thompson et al. 2011). Patterns in sunspot magnetic field strengths and helioseismological measurements have been reported suggesting that there may be stranger behavior to come (Penn and Livingston~2006, 2011, Hill et al.~2011).

Polar fields are believed to be largely determined by the poleward transport of the remnant magnetic flux of decayed active regions, as we discuss below. Attempts have been made to explain the weak polar fields of Cycle 23 using flux-transport dynamo models and surface-flux-transport models. Dikpati~(2011) argued from a simple numerical estimate and from detailed flux-transport dynamo modeling that the weakness of the Cycle 23 polar fields may be mostly explained by the relative weakness of the active-region fields during Cycle 23. On the other hand, Jiang et al.~(2011) found that the weak polar fields of Cycle 23 could be reproduced by a major increase, 55\%, in the meridional flow during that cycle. They could also be reproduced by a decrease in the average tilt angle of sunspots but this would also lead to a 1.5-year delay of the Cycle 23 polar field reversal that was not observed.

Measurements of the polar magnetic field are extremely challenging because of the foreshortening of the solar surface near the limb as well as the intrinsic weakness of the fields near the poles. Yet these measurements are critical because the polar fields play a crucial role in determining the morphology of the global coronal magnetic field. Global models of the coronal field, such as PFSS and MHD models, rely on full-surface maps of the photospheric field as boundary conditions and are particularly sensitive to the polar fields, especially near solar minimum.

Following the discovery by Babcock~(1959) that the solar polar field reverses with the sunspot cycle, Babcock~(1961) produced a powerful phenomenological description of the solar cycle based on his magnetograph observations. This model was then developed by Leighton~(1964, 1969) into a kinematic model for the transport of photospheric magnetic flux. This model was related to Parker's~(1955) idea of the ``$\alpha$ effect''. Differential rotation in the solar convection zone produces from a poloidal field a significant toroidal field component. These can become unstable and emerge buoyantly into the atmosphere to produce bipolar active regions. During their buoyant rise, the Coriolis force tilts these regions with respect to the equator such that, at the beginning of each activity cycle, the emerging regions generally have poloidal field component opposing the poloidal field associated with the two magnetic polar caps. The new poloidal active-region fields associated with the emerging active regions thereby oppose the polar fields until they cancel them and reverse their polarity. This model, the ``Babcock-Leighton'' model has been modified by Wang \& Sheeley~(1991) and Wang et al.~(1991) to relate the polar flux distribution to the observed poleward meridional transport of decayed active-region flux. Pole-ward meridional flows of 20~m/s were seen in Mt. Wilson (Ulrich~1993) and GONG (Hathaway et al.~1996) Doppler measurements. The polar fields are also believed to supply the seed field for the generation of the next activity cycle, being transported equatorward by a subsurface return meridional flow.

In this paper we seek correlations between active region and polar fields in long, continuous series of magnetograph observations. We analyze 37 years of full-disk observations from 3 NSO Kitt Peak magnetographs that have been observing daily, weather permitting, since 1974: the Synoptic Optical Long-term Investigations of the Sun (SOLIS) Vector-Spectromagnetograph (VSM), that has been observing the 630.2~nm Fe~{\sc I} spectral line since August 2003 (Keller et al.~2003); the Spectromagnetograph (SPMG), that observed the 868.8~nm  Fe~{\sc I} spectral line from November 1992 until September 2003 (Jones et al.~1992); and the 512-channel instrument, that observed the 868.8~nm  Fe~{\sc I} spectral line from February 1974 until April 1993 (Livingston et al.~1976). Together the data set covers Cycles 21-23 in their entirety, as well as the end of Cycle 20 and the beginning of Cycle 24, the current cycle, with $1^{\prime\prime}$ pixel$^{-1}$ spatial resolution. The Mount Wilson 150-foot tower magnetograph (Howard~1976) has been observing the full disk in the Fe~{\sc I} line at 525.0~nm also since 1974. These observations are taken on a non-uniform grid of lower spatial resolution, $12^{\prime\prime}$, and are then interpolated onto a higher-resolution regular grid with pixel size approximately $4^{\prime\prime}$. The raw materials of the project are magnetograms of the full solar disk in sky coordinates. We will refer to these magnetograms as ``full-disk'' images or simply ``images''. The two data sets from Kitt Peak and Mt. Wilson give us an opportunity to study the evolving relationship between the active region and polar fields in some detail over several solar cycles. Tlatov et al.~(2010) analyzed longitudinal magnetograms from Kitt Peak (1975-2002) and Michelson Doppler Imager on board the Solar and Heliospheric Observatory (MDI/SOHO, 1996-2009) to study the latitude distribution and tilt of magnetic bipoles during Cycles 21-23. They found distinctly different behavior for small quiet-Sun bipoles, ephemeral regions, and active regions. Here we will simply report the average behavior of active region fields and compare these fields with fields at high latitudes.

A major concern with studies of magnetic field measurements from multiple telescopes is that the observed field strength depends on many details of the observing system (e.g., Ulrich et al.~2009). Numerous studies have shown that there can be significant systematic differences between data taken by different telescopes at the same time (e.g., Jones et al.~1993, Walton et al.~1993, Jones and Ceja~2001, Thornton and Jones~2002, Wentzler et al.~2004, Demidov et al.~2008). For a good measurement of the polar field strengths a well-determined zero point is essential. For example, the work by Arge et al.~(2002) required a zero-point correction to be applied to some of the early NSO data. Here we use SPMG and 512-channel data that have been recently 
cleaned by Jack Harvey, including a zero-point correction without which the polar fields would not be reliable. The aperture size of the Mt. Wilson instrument has changed over time but here we limit the study to data from the $12^{\prime\prime}$ configuration, so this does not affect the Mt Wilson data series analyzed here.



In this paper we use the term ``field strength'' to denote pixel-averaged magnetic flux density in units of Gauss (Mx~cm$^{-2}$). This applies to both the line-of-sight measurements produced by the Kitt Peak and Mt. Wilson observatories and the derived radial data. This type of data set is distinct from direct measurements of the full magnetic vector field strength described by, e.g., Penn \& Livingston~(2011) and Pevtsov et al.~(2011).

The paper is organized as follows. We begin by summarizing the data using a butterfly diagram in Section~\ref {s:butterfly}. We investigate the active region field properties in more detail in Section~\ref{s:activeregion}. We then compare the active region fields to the high-latitude stream fields and the polar field changes in Section~\ref{s:arstpl} and then describe a possible explanation for the weakness of the polar fields since the maximum of solar Cycle 23. In Section~\ref{s:chromosphere} we analyze the SPMG, SOLIS and Mt. Wilson chromospheric and high photospheric longitudinal field measurements that provide full-disk coverage from the beginning of Cycle 23 to the present. We conclude in Section~\ref{s:conclusion}.

\section{Butterfly diagrams}
\label{s:butterfly}

\begin{figure} 
\begin{center}
\resizebox{0.99\textwidth}{!}{\includegraphics*[20,255][600,530]{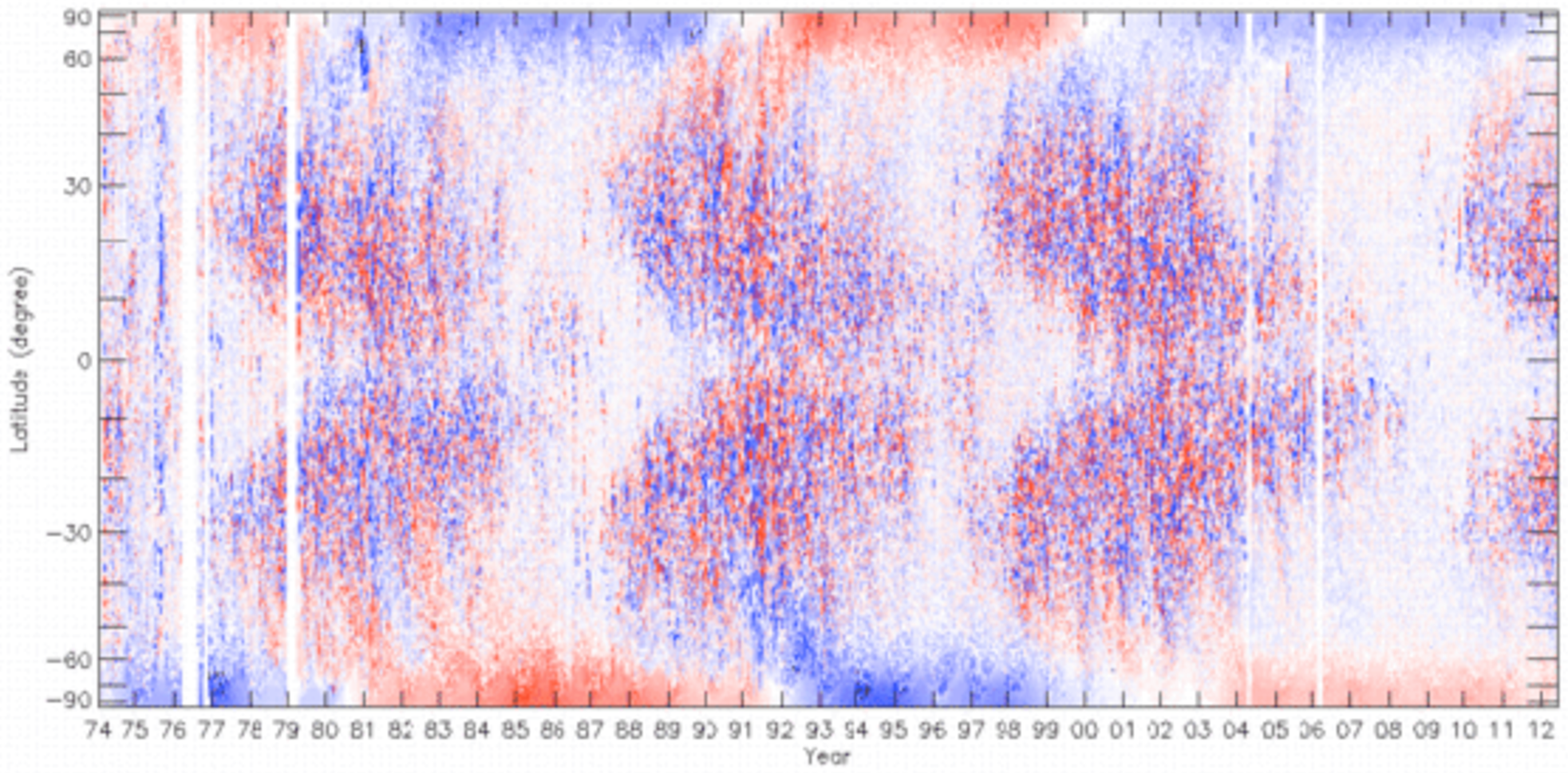}}
\resizebox{0.99\textwidth}{!}{\includegraphics*[20,255][600,530]{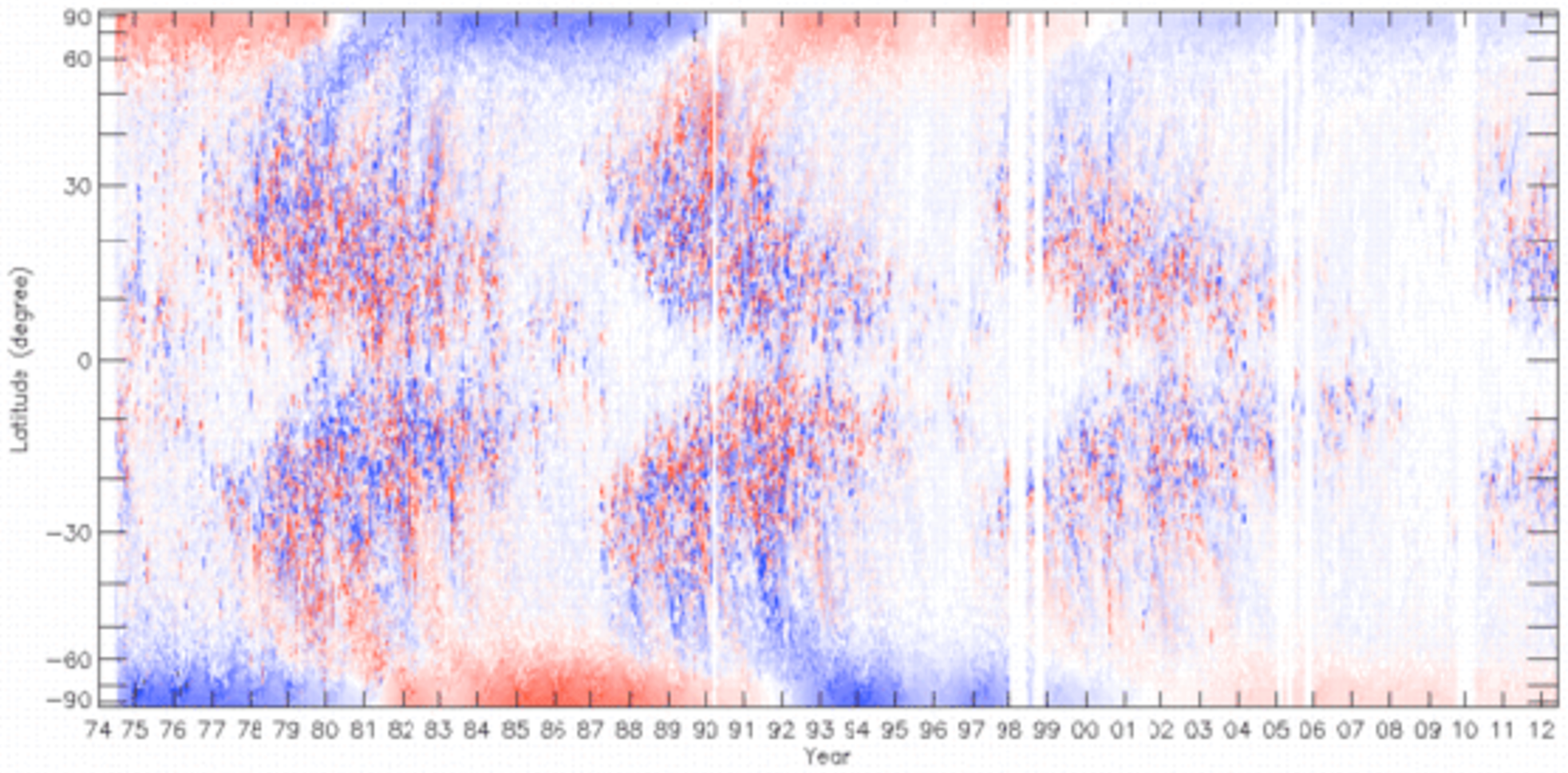}}
\end{center}
\caption{Butterfly diagrams based on Kitt Peak (top) and Mt. Wilson (bottom) data, summarizing the photospheric radial field distributions derived from the longitudinal photospheric field measurements of the two observatories. Each pixel is colored to represent the average field strength at each time and latitude. Red/blue represents positive/negative flux, with the color scale saturated at $\pm 20$~G.}
\label{fig:butterfly}
\end{figure}

The butterfly diagrams shown in Figure~\ref{fig:butterfly}, one for each observatory, are formed by binning all pixels (with central meridian distance $30^{\circ}$ or less) from each image into 180 equal-size bins in sine(latitude) and taking the average of each bin. This process produces a 180-element array in sine(latitude) for each image, and we combine these along a time axis to form the two-dimensional space-time maps shown in the figure. The radial field component is shown, derived from the longitudinal measurements by dividing by the cosine of the heliocentric angle $\rho$ (the angle between the line of sight and the local solar radial vector). The time axis spans the approximately 37 years between 1974 and the early months of 2012, covering three full solar cycles, 21-23, as well as the end of Cycle 20 and the beginning of Cycle 24. In the construction of the butterfly diagrams, two corrections have been applied: the longitudinal field measurements have been turned into data for the radial field component, and poorly observed and unobserved fields near the pole have been estimated to provide data for all latitudes at all times. We describe these two corrections before discussing the butterfly diagrams in detail.

\subsection{Radial field correction}
\label{s:radialcorr}

One 	complication of the study is the derivation of the radial photospheric field component from longitudinal measurements. In the past, the inclination of photospheric fields has been diagnosed in several ways. Using large time-series of full-disk images from Stanford's Wilcox Solar Observatory (WSO), Svalgaard et al.~(1978) calculated the average line-of-sight field strength across the disk within an equatorial band and found that it varied as $\cos (\rho )$, a result consistent with a radial photospheric field (see also Rudenko~2004). Howard~(1974,~1991) inferred the azimuthal component of the photospheric field by comparing the magnetic flux of a region when it was located at equal distances east and west of the central meridian, and found that azimuthal (east/west) tilts of photospheric fields were generally less than $10^{\circ}$. Topka et al.~(1992) drew a similar conclusion from the center-to-limb variations of continuum contrasts of facular areas.  These findings are in line with physical expectations. Petrie and  Patrikeeva~(2009) forward-fitted the longitudinal components of tilted vectors to disk-passage flux profiles using SOLIS photospheric and chromospheric magnetograms and found that most photospheric fields are tilted at angles of $12^{\circ}$ or less whereas chromospheric fields have a much broader distribution of tilt angles. The photospheric magnetic field is dynamically dominated by dense plasma and is confined by the ram pressure of supergranular flows to network boundaries.  The resulting intense flux tubes are much more buoyant than their surroundings and their near-vertical orientation is believed to be due to this buoyancy (Parker~1955,~1966). During the past several years, the SOT-SP instrument on Hinode has been periodically observing vector fields of the polar regions, showing many vertically oriented intense flux tubes as well as ubiquitous horizontal fields (Tsuneta et al.~2008, Shiota et al.~2012). Petrie \& Patrikeeva~(2009) presented evidence that the polar photospheric fields are approximately radial. Moreover, it has been known for a long time that potential field models of the corona more successfully reproduce observed coronal structures if they have been extrapolated from derived radial photospheric boundary data than if line-of-sight measurements have been directly applied as boundary conditions (Wang \& Sheeley~1992). These results justify the derivation of radial photospheric field maps from longitudinal photospheric field measurements.

\subsection{Polar field correction}
\label{s:polarcorr}

An important advantage of studying approximately radial fields is that the full magnetic flux through the photosphere can be estimated reasonably accurately over most of the solar disk. However, because the solar rotation axis is tilted at an angle of $7.25^{\circ}$ with respect to the ecliptic plane, the fields near the solar poles are either observed with very large viewing angles or not observed at all for six months at a time. Also the noise level is inflated near the poles by the radial field correction described in Subsection~\ref{s:radialcorr} above. For these reasons we have had to fill the locations in the butterfly diagram nearest the poles using estimated values for these fields. This is a well-known problem in the construction of synoptic magnetograms. We employ techniques similar to those described by Sun et al.~(2011). Annual averages of all high-latitude fields were calculated from measurements taken with advantageous solar axis tilt ($B_0$) angles ($B_0>5^{\circ}$ for northern and $B_0<5^{\circ}$ for southern high-latitude fields). At each latitude a cubic spline interpolation of these annual average measurements was calculated on the original time grid. The measurements of the high-latitude fields are only available for a range of latitudes and this range is strongly dependent on the $B_0$ angle. The annual averages were well defined and regular functions of time for all but the two most poleward sin(latitude) bins at each pole. To estimate the values for the two unreliable bins at each pole, a one-dimensional quintic polynomial fit was calculated for each image, based on a latitude- and $B_0$-angle-dependent weighted average of the available measurements and the spline fits, assuming symmetry about the pole. The resulting simulated data were only used for apparent latitude (the heliographic latitude with the $B_0$ angle subtracted) greater than $75^{\circ}$ or less than $-75^{\circ}$. For apparent latitudes between $\pm60^{\circ}$ and $\pm75^{\circ}$ a linear combination of the real and simulated measurements was used.

\subsection{Discussion of the butterfly diagrams}
\label{s:butterflydisc}

Figure~\ref{fig:butterfly} shows several distinctive patterns in the long-term behavior of the fields. The active fields begin each cycle emerging at latitudes around $\pm 30^{\circ}$ and subsequently emerge at progressively lower latitudes on average, creating the distinctive wings of the butterfly patterns. The diagrams also shows the change of polarity of the polar fields once each cycle, and the coincidence of each polar-field polarity change with the activity maximum of its cycle. Between the active and polar latitudes there is clear evidence of poleward flux transport, which appears most intense at active phases of the cycle and at times of polar field change. Most of the flux that emerges in active regions cancels with flux of opposite polarity but a proportion survives as weak flux that is carried poleward by the meridional flow. This poleward drift of the weak, decayed magnetic flux appears as plumes of one dominant polarity at high latitudes, between about $40^{\circ}$ and $65^{\circ}$. Later in this paper we will explore correlations between these three classes of magnetic field: active-region, polar, and high-latitude stream fields.

The recent Cycle 23 minimum was unusually long and quiet, as Figure~\ref{fig:butterfly} shows. Comparing the fields around 2008-2010 with those around 1985-87 and 1996-97, the recent minimum includes a two-year interval during which almost no activity appears in Figure~\ref{fig:butterfly}, in contrast to the preceding two minima. In the Cycle 23 minimum there were intervals several Carrington rotations long during which no significant active fields appeared (e.g., CR~2082). During the previous two minima a rotation didn't pass without some activity appearing on the disk. Cycle 23's polar fields also stand out as being weaker than the polar fields for any other cycle. The active region fields of Cycle 23 also appear weaker than those of the preceding two cycles in the Mt. Wilson plot in Figure~\ref{fig:butterfly} but this is not clear in the Kitt Peak plot. According to both plots the two hemispheres were approximately symmetric until the descending phase of Cycle 23. Since around 2005 there has been major asymmetry between the two hemispheres. During the descent into the Cycle 23 minimum the southern hemisphere remained active a year or two after the northern hemisphere had gone quiet. After the very long and quiet Cycle 23 minimum, Cycle-24 fields appeared first in the Northern hemisphere and, although both hemispheres showed increased activity in early 2010, the active regions in the north have been both stronger and more numerous than those in the south.

\begin{figure} 
\begin{center}
\resizebox{0.9\textwidth}{!}{\includegraphics*[10,230][600,530]{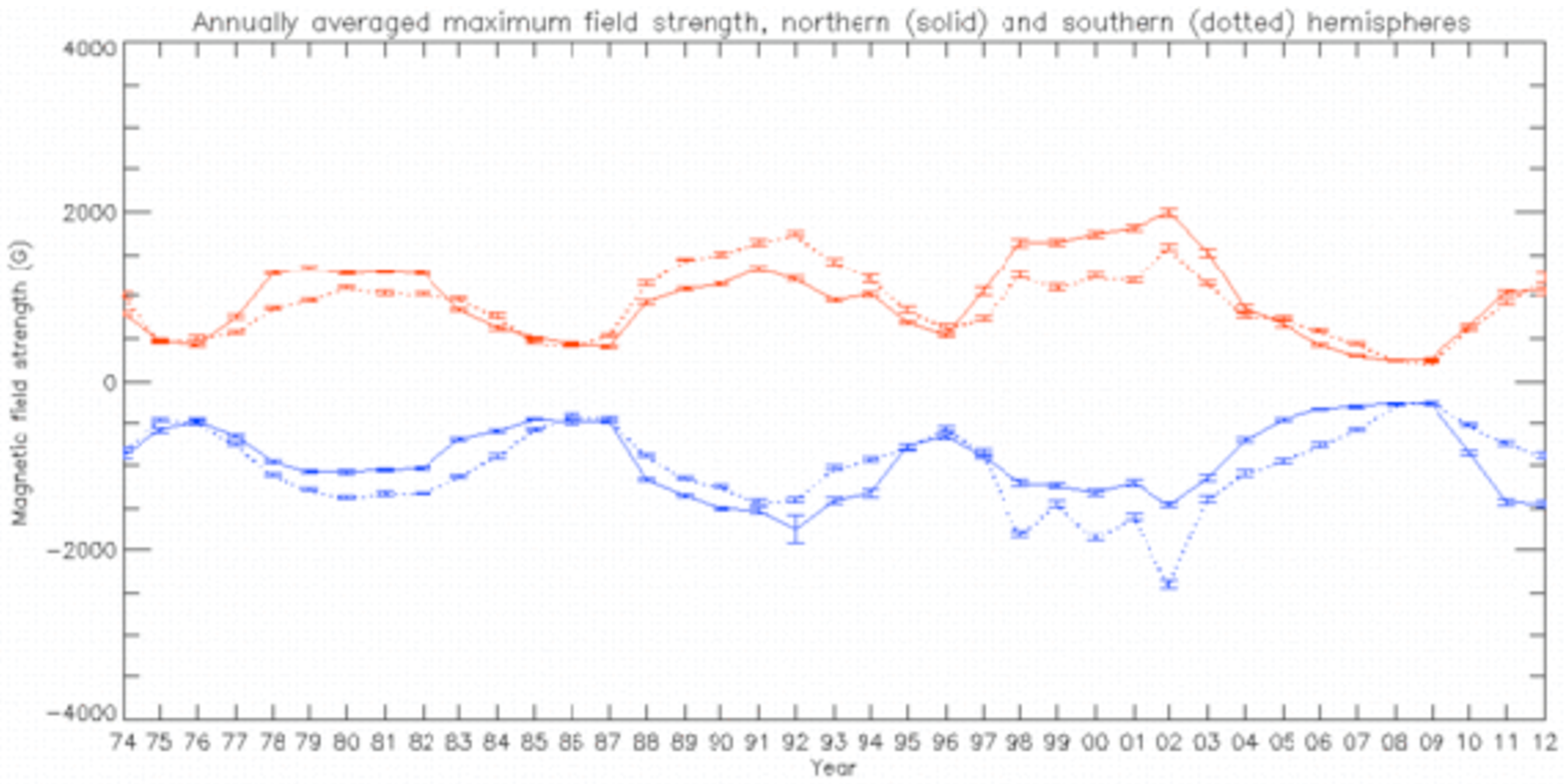}}
\resizebox{0.9\textwidth}{!}{\includegraphics*[10,230][600,530]{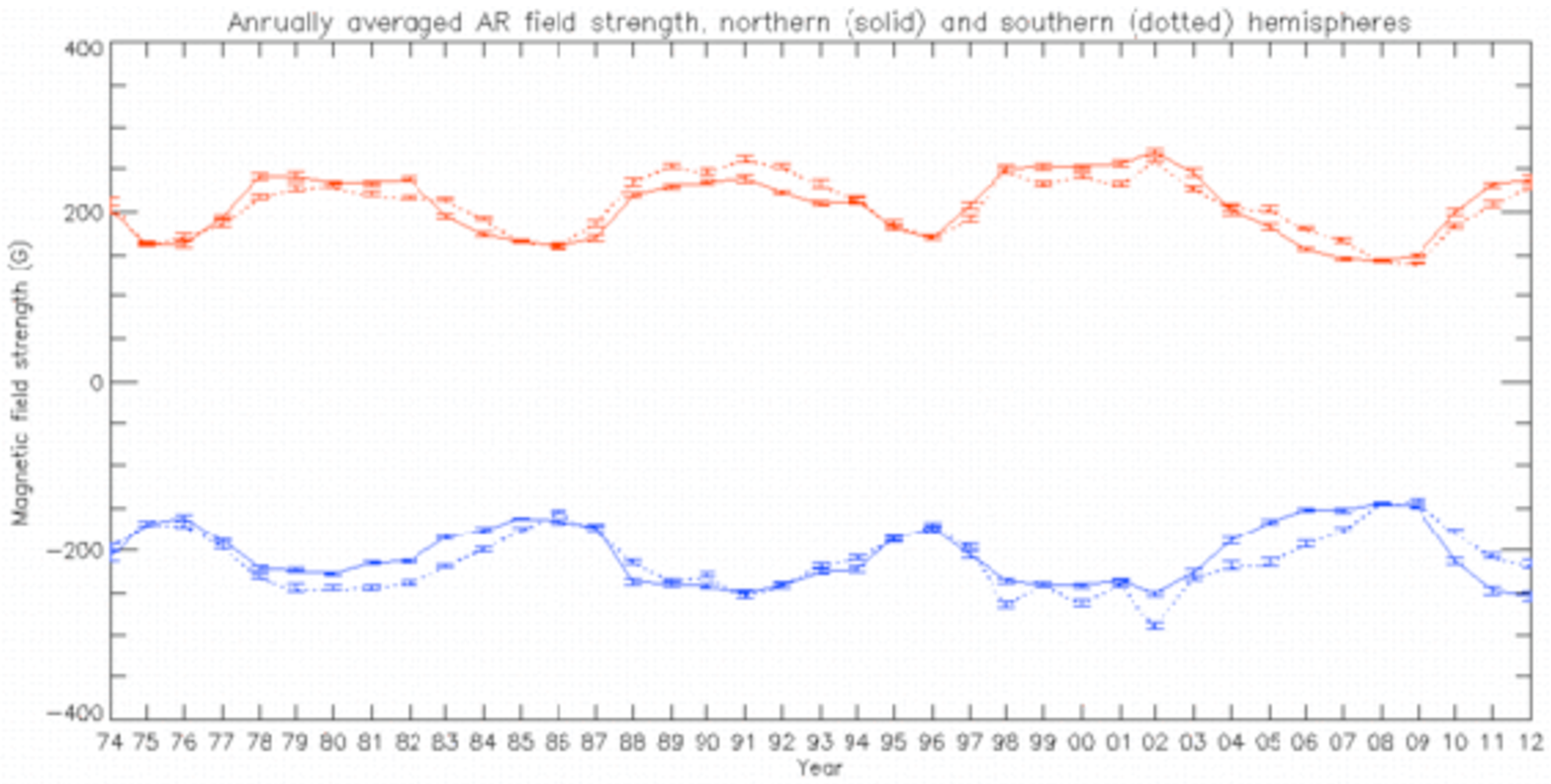}}
\resizebox{0.9\textwidth}{!}{\includegraphics*[10,230][600,530]{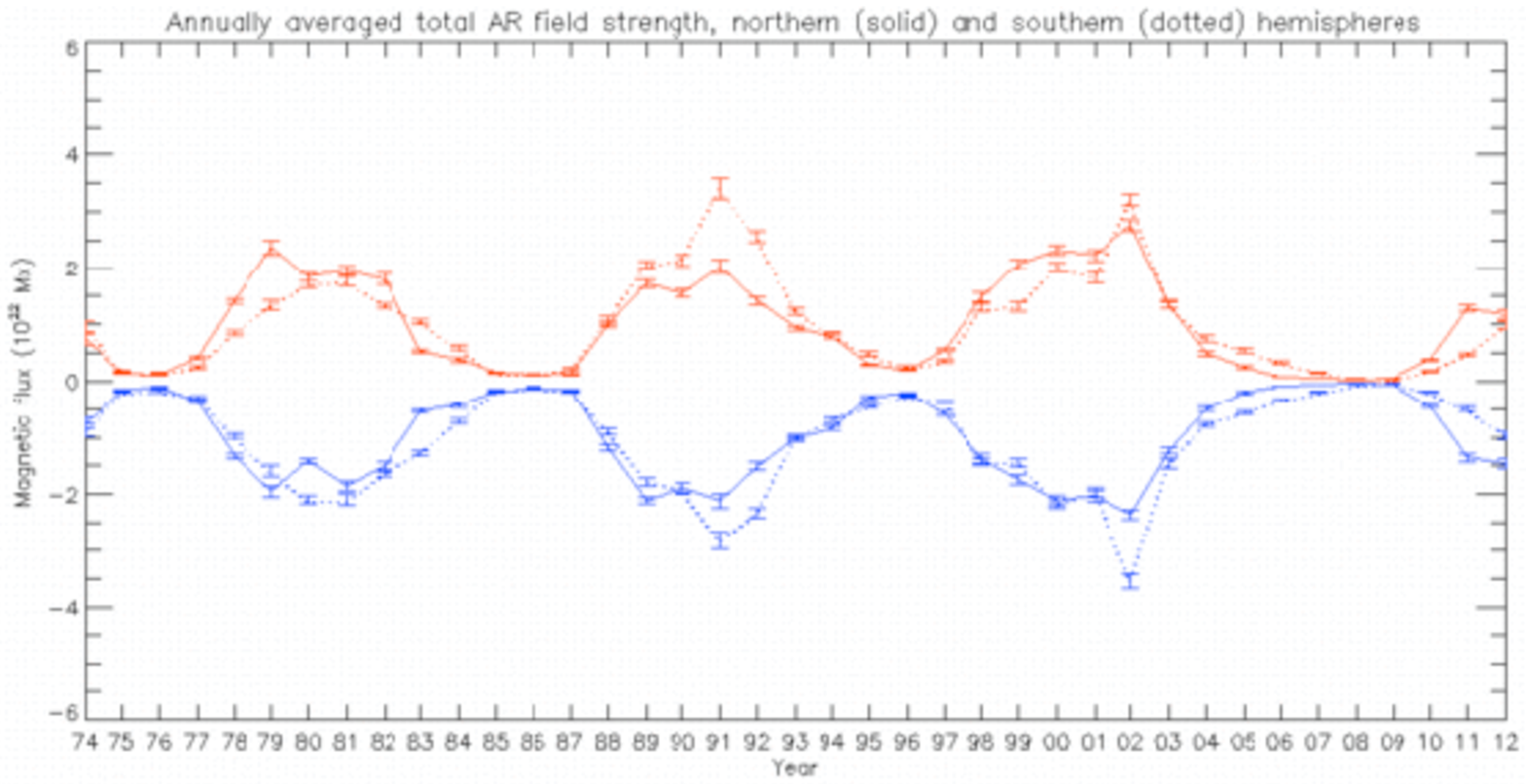}}
\end{center}
\caption{The Kitt Peak year-averaged maximum active-region field strength (top), average active-region field strength (middle) and total active-region flux (bottom), plotted for northern (solid lines) and southern (dotted lines) hemispheres and for positive (red) and negative (blue) polarity. The error bars indicate the standard deviations of the annual means.}
\label{fig:arflux}
\end{figure}

\begin{figure} 
\begin{center}
\resizebox{0.9\textwidth}{!}{\includegraphics*[10,230][600,530]{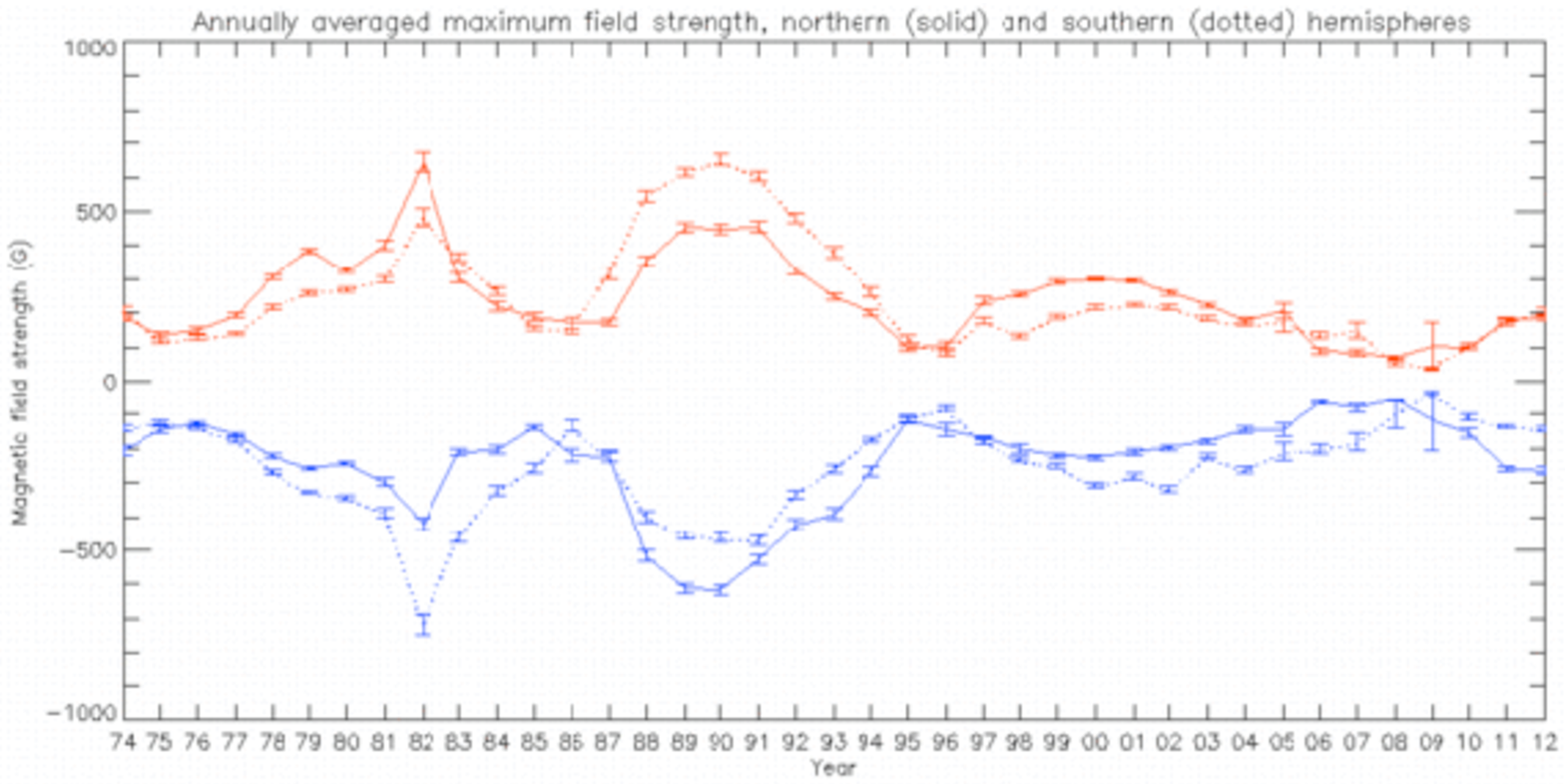}}
\resizebox{0.9\textwidth}{!}{\includegraphics*[10,230][600,530]{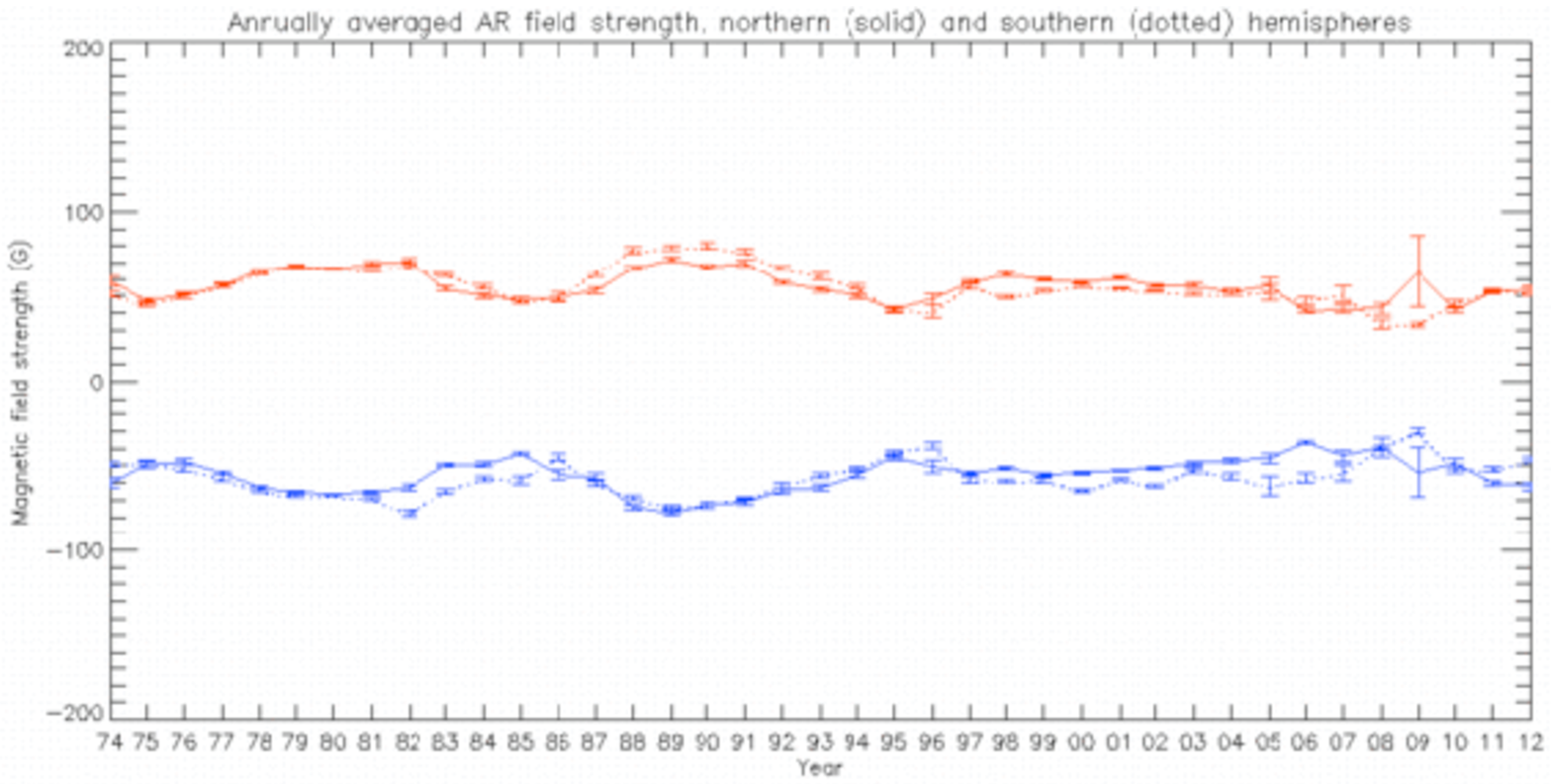}}
\resizebox{0.9\textwidth}{!}{\includegraphics*[10,230][600,530]{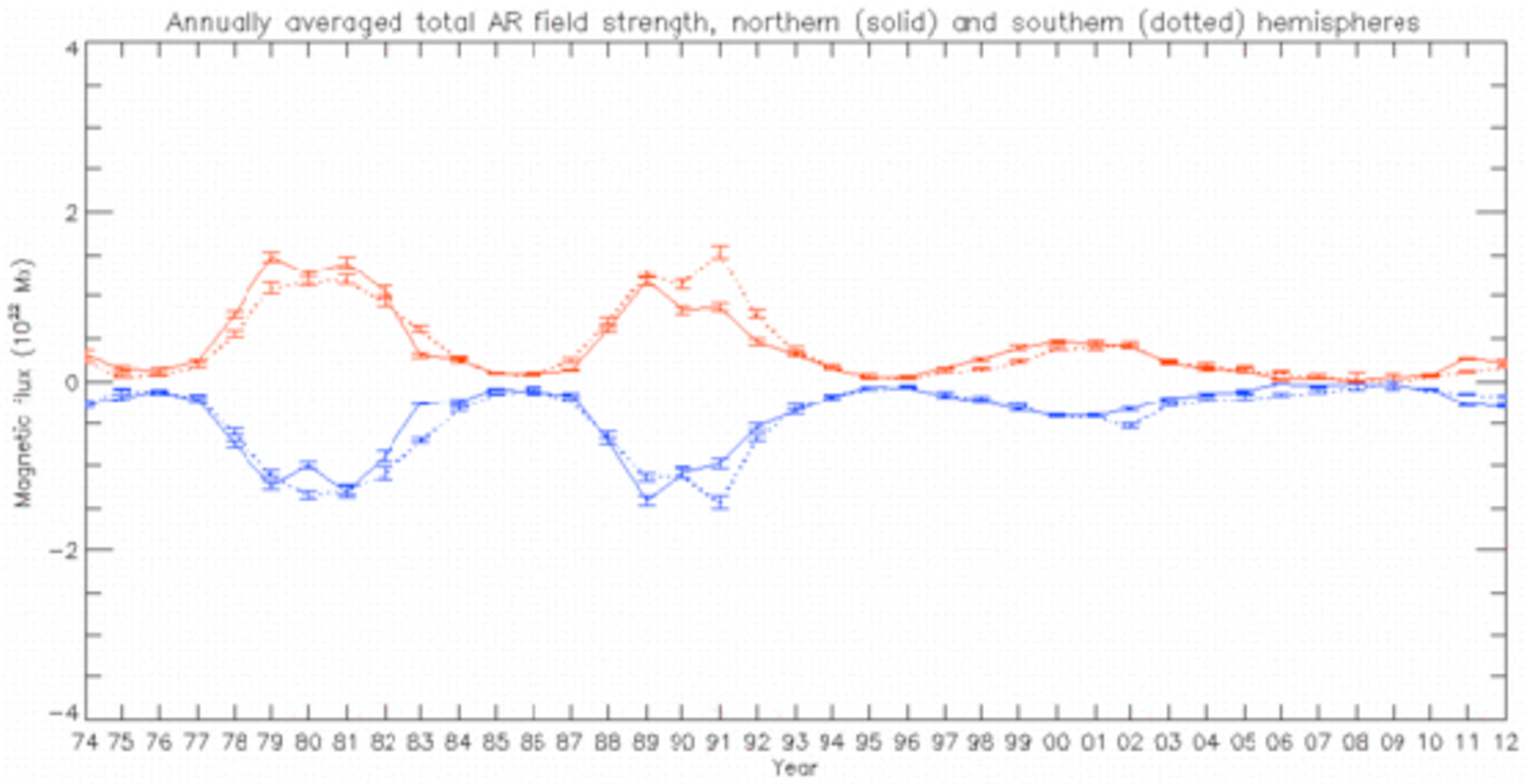}}
\end{center}
\caption{The Mt. Wilson year-averaged maximum active-region field strength (top), average active-region field strength (middle) and total active-region flux (bottom), plotted for northern (solid lines) and southern (dotted lines) hemispheres and for positive (red) and negative (blue) polarity. The error bars indicate the standard deviations of the annual means.}
\label{fig:arflux_mwo}
\end{figure}

\begin{figure} 
\begin{center}
\resizebox{0.9\textwidth}{!}{\includegraphics*[10,230][600,530]{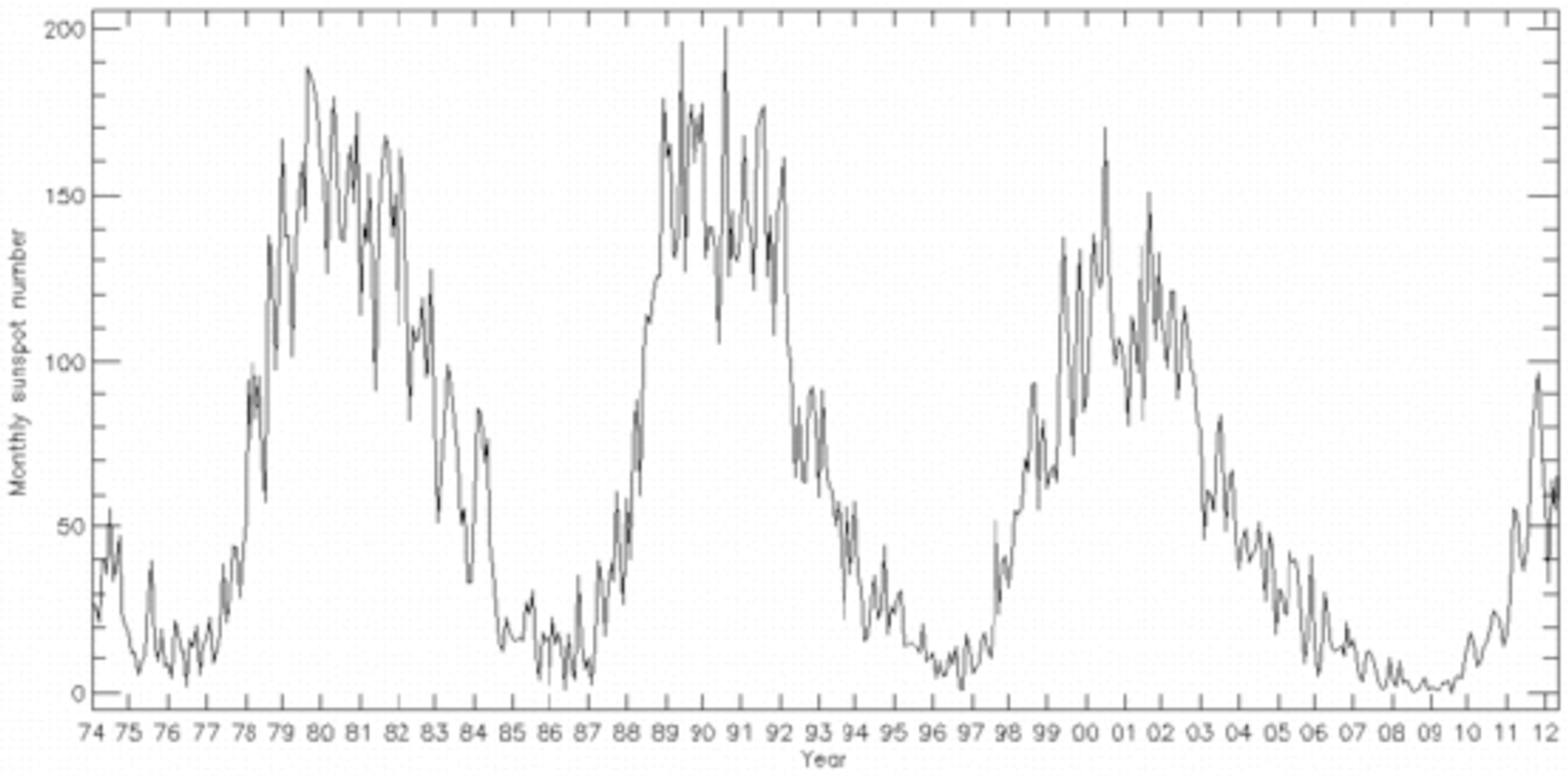}}
\end{center}
\caption{The monthly sunspot number from 1974 to 2012. Comparing with the bottom panels of Figures~\ref{fig:arflux} and \ref{fig:arflux_mwo}, the sunspot number shows more variation between the peaks of the cycles than the Kitt Peak magnetic flux measurements and less than the Mt. Wilson magnetic flux measurements.}
\label{fig:ssn74-12}
\end{figure}

\section{The active region fields}
\label{s:activeregion}

\subsection{The average field strengths}
\label{s:arfieldstrengths}

The active region fields are identified by field strength. Pixels of absolute value 100~G or stronger in the Kitt Peak images are deemed to represent active region fields. The Mt. Wilson images are of significantly lower spatial resolution and a threshold of 25~G was found to identify approximately the same fields. A range of alternative thresholds was tried and the results were qualitatively unchanged.

Figures~\ref{fig:arflux} and \ref{fig:arflux_mwo} show the year-averaged maximum active-region field strength, average active-region field strength and total active-region flux, plotted for north and south hemispheres and for positive and negative field polarity. These averages were derived by taking the per-image maximum active region field strengths, average active region field strengths and total active region fluxes, for each hemisphere and for each field polarity, and computing their annual averages. In these plots as in the butterfly plots of Figure~\ref{fig:butterfly} the Cycle 23 active region fields are significantly weaker than those of previous cycles in the Mt. Wilson data but not in the Kitt Peak data. This, like the corresponding differences in the butterfly diagram in Figure~\ref{fig:butterfly}, seems likely to be mostly because of instrument changes at Kitt Peak. Comparing Figures~\ref{fig:arflux} and \ref{fig:arflux_mwo} it also becomes clear that the fields and fluxes are generally significantly weaker in the Mt. Wilson data than in the Kitt Peak data. This is most likely because the Mt. Wilson pixels size is approximately four times larger (in each dimension) than the Kitt Peak pixel size of $1^{\prime\prime}$.

The plots of the maximum and average fields show the expected hemispheric asymmetry during cycles 21-23. Throughout Cycle 21 northern/southern active regions had stronger positive/negative than negative/positive fields on average. During Cycle 22 the opposite was true and the pattern reversed again when Cycle 23 began. But after about 2003-04 the active-region asymmetry pattern is overwhelmed by the hemispheric asymmetry  in the plots.  During the descending phase of Cycle 23 the southern fields of both polarities are stronger than the northern fields, and since the beginning of Cycle 24 the northern fields of both polarities have been stronger than the southern fields. However, the maximum and average field strengths have remained asymmetrically distributed between the polarities in each hemisphere since the maximum of Cycle 23: before the cycle 23 minimum northern/southern active regions had stronger positive/negative than negative/positive fields on average, and this pattern has faithfully reversed for Cycle 24.

The plots of the total active region flux in Figures~\ref{fig:arflux} and \ref{fig:arflux_mwo} (bottom panels) do not show such alternating patterns. According to each observatory, the total flux is approximately balanced within each hemisphere, giving these plots a symmetric appearance. The three cycle minima captured in the data set are distinctly different in character. The Cycle 22 minimum was the simplest of the three. Both hemispheres descended into the minimum at approximately the same rate and the same time and promptly ascended into Cycle 23 in a similarly symmetric fashion. The descents into the Cycle 21 and 23 minima were both lead by the southern hemisphere and this hemisphere lagged the north in ascending to Cycles 22 and 24. The Cycle 23-24 transition was much longer than the Cycle 21-22 transition, and  the north-south asymmetry significantly greater at the beginning of Cycle 24 than at the beginning of Cycle 22, but the north-south asymmetry during the descending phase of Cycle 21 was almost as strong as during the descending phase of Cycle 23.

The bottom panels of Figures~\ref{fig:arflux} and \ref{fig:arflux_mwo} show that the activity level has been rising steadily since 2009 but it is still 50\% or less than the strengths of the activity maxima of Cycles 21-23 according to the annually averaged total flux measurements. In terms of the annual maximum field strengths (top panel) the 2011 level in the northern hemisphere is only slightly less than the maximum levels of Cycles 21-23 but the southern hemisphere is still less than 50\% of the maximum levels of Cycles 21-23. The total flux detected during the individual cycles varies greatly between the two observatories, with Mt. Wilson detecting significantly less than Kitt Peak during Cycles 22 and (especially) 23. This can be explained by the better spatial resolution of the Kitt Peak magnetographs. Whereas Kitt Peak measured comparable quantities of flux during each of the three full cycles, the Cycle 23 total active region flux is much less than for previous cycles in the Mt. Wilson data. This discrepancy is partly explained by the three Kitt Peak magnetographs' differing ability to measure strong fields values. For example, the saturation values for the SPMG and SOLIS/VSM instruments that observed during Cycles~23 and 24 are significantly higher than for the 512-channel instrument that observed during Cycles~21 and 22. On the other hand, it has been suggested that the strong temperature sensitivity of the Fe~~{\sc I} 525.0~nm line used at Mt. Wilson, and blends of this line in sunspots, may complicate measurements in this line (Socas-Navarro et al.~2008). Figure~\ref{fig:ssn74-12} shows a plot of the monthly international sunspot number\footnote{http://www.ngdc.noaa.gov/stp/solar/ssndata.html} for comparison with Figures~\ref{fig:arflux} and \ref{fig:arflux_mwo}. The sunspot number shows a decrease in the amplitude of Cycle 23 compared to Cycles 21 and 23, unlike Figures~\ref{fig:arflux}, but not as strong as the decrease shown in Figure~\ref{fig:arflux_mwo}. The real behavior seems to follow a trend somewhere in between those shown in Figures~\ref{fig:arflux} and \ref{fig:arflux_mwo}. Because of these complications in the data, we need to be careful not to draw conclusions dependent on idiosyncrasies of a particular data set. This is one reason why it is important to analyze the data from the two observatories separately.

Penn \& Livingston~(2006, 2011) have measured the full magnetic field vector strength by the separation between the two Zeeman components of a magnetically sensitive 1564.8~nm Fe~{\sc I} spectral line at the NSO Kitt Peak McMath-Pierce telescope over about a decade. They have reported a steady decreasing trend in annual average sunspot field strengths, from over 2500~G in 2002 to about 2000~G in 2011, accompanied by steady increase in umbral intensity. Their observations differ from ours in important ways, including their selection criterion being the darkest part of the sunspot umbrae and the fact that they directly measure the strength of the full magnetic vector where we work with pixel-averaged measurements of the longitudinal field component. Pevtsov et al.~(2011) analyzed historic synoptic sunspot magnetic field strength measurements taken in a similar way to Penn \& Livingston's measurements at seven observatories in the former USSR covering the period from 1957 to 2011, mostly observed using the 630.2 nm Fe~{\sc I} spectral line although other lines were used as well. They found that, while sunspot field strengths showed expected cyclical variations, there was no evidence of a gradual decrease in sunspot magnetic fields over the four and a half solar cycles covered by their data. They found that the minimum of Cycle~23 was the weakest in their data set (which included Cycles~19-23 and the beginning of Cycle~24) but that the sunspot fields have been becoming steadily stronger since 2008. This pattern is similar to the one that we see in the top panels of Figures~\ref{fig:arflux} and \ref{fig:arflux_mwo}. Watson et al.~(2011) analyzed pixel-averaged longitudinal magnetic field measurements from SoHO/MDI magnetogram data covering Cycle 23 and found only a minor decreasing trend in maximum umbral magnetic fields. The direct measurements of sunspot field strengths described by Penn \& Livingston~(2011) and Pevtsov et al.~(2011) are less sensitive than magnetogram measurements to scattered light but this fact does not explain the discrepancies between the above results.

\begin{figure} 
\begin{center}
\resizebox{0.99\textwidth}{!}{\includegraphics*[20,255][600,530]{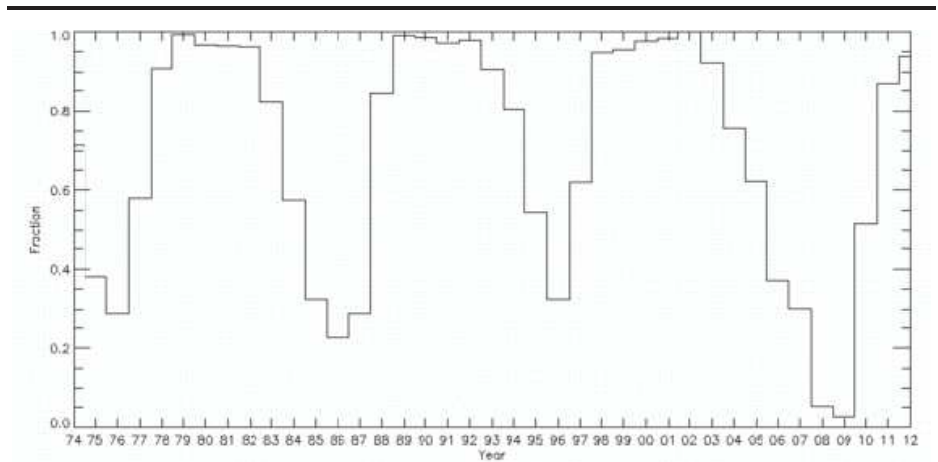}}
\end{center}
\caption{A histogram of the proportion of Kitt Peak full-disk images that included fields stronger than 1 kG during each year.}
\label{fig:kgnormhist}
\end{figure}

Figure~\ref{fig:kgnormhist} shows a histogram of the proportion of full-disk images that included fields stronger than 1 kG during each year. The relative quietness of the Cycle 23 minimum compared to the two previous minima is clear in the plot. During 2008 and 2009 fewer than 5\% of the images contained kG fields whereas during the Cycle 21 and 22 minima the ratio never dipped below 20\% and 30\%, respectively. Since 2009 the proportion of kG fields has risen steadily, to about 55\% in 2010, over 85\% in 2011 and nearly 95\% so far in 2012. These results emphasize the depth of the Cycle 23 minimum and show that the increase in activity from 2010 onward has been impressive. According to some (but not all) indices, conditions are approaching the levels of past activity maxima at the time of writing.



\begin{figure} 
\begin{center}
\resizebox{0.9\textwidth}{!}{\includegraphics*[10,230][600,530]{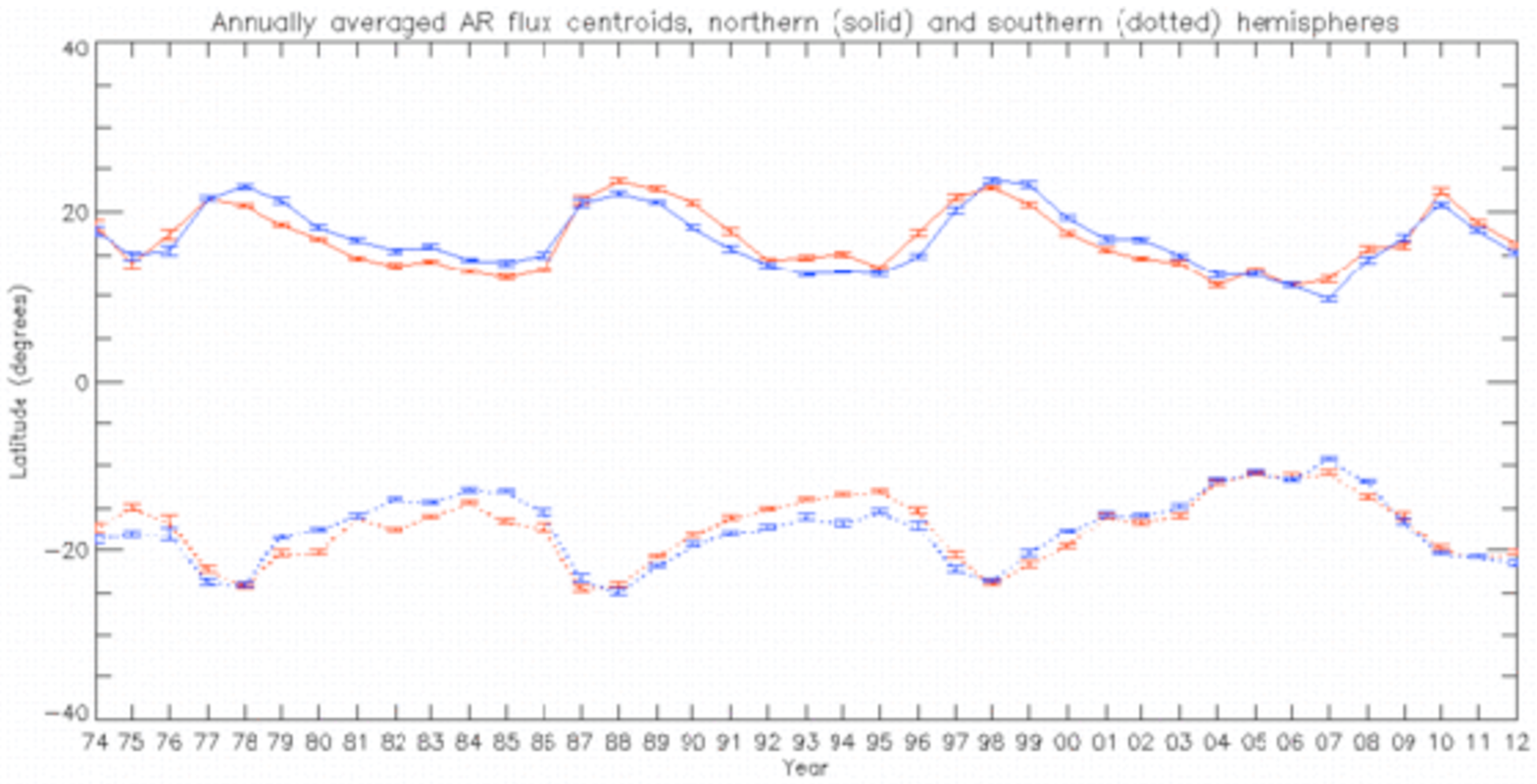}}
\resizebox{0.9\textwidth}{!}{\includegraphics*[10,230][600,530]{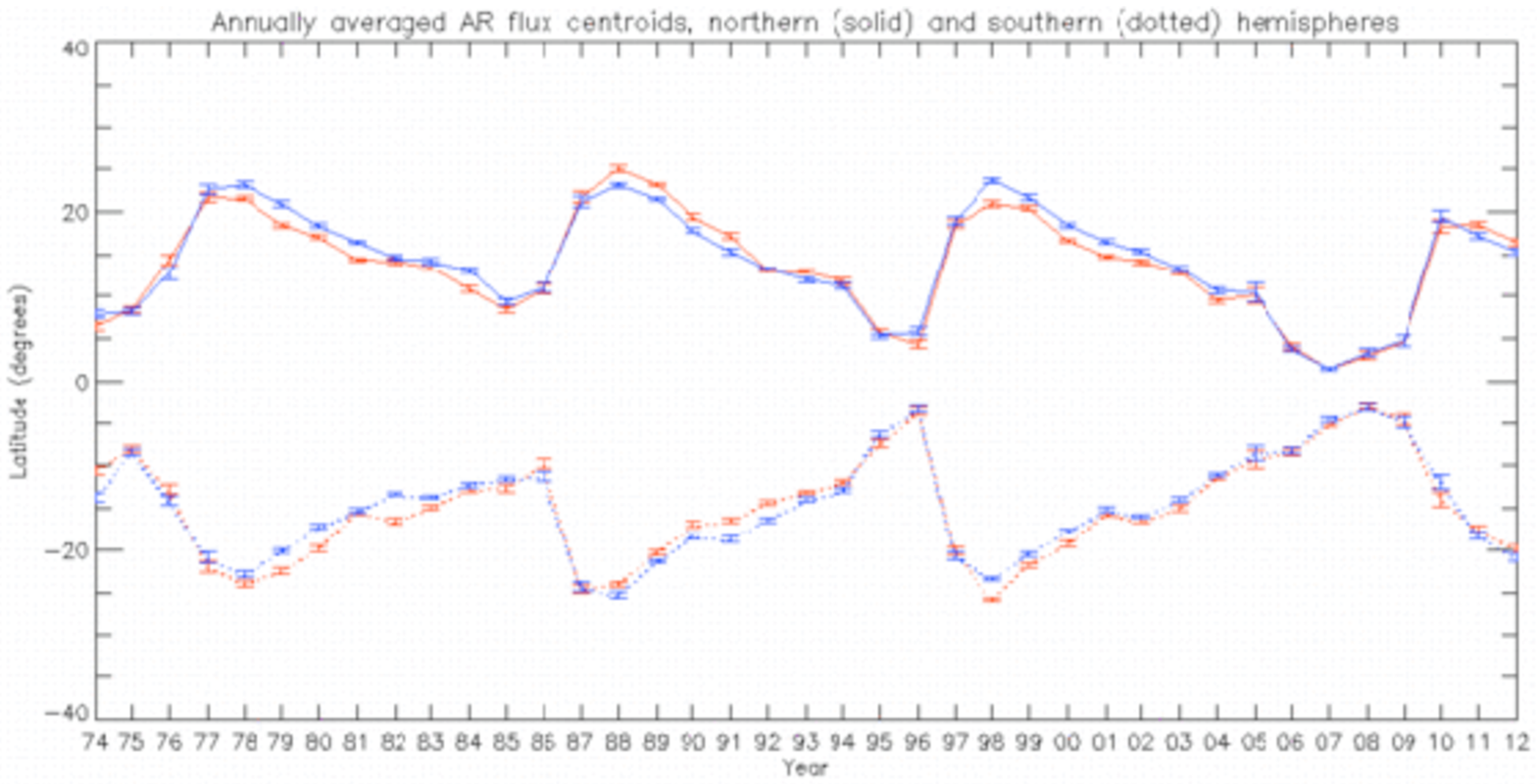}}
\end{center}
\caption{The latitudinal centroids of the active fields as functions of time for positive (red) and negative (blue) polarities in the northern (solid lines) and southern (dotted lines) hemispheres for the Kitt Peak data (top) and the Mt. Wilson data (bottom).  The error bars indicate the standard deviations of the annual means.}
\label{fig:arcentroidave}
\end{figure}

\section{The hemispheric active region latitude centroids}
\label{s:arcentroids}

To examine further the active field asymmetries we plot the latitudinal centroids of the active fields as functions of time in Figure~\ref{fig:arcentroidave} for both observatories. These centroids were calculated image-by-image for each polarity and for each hemisphere. Figure~\ref{fig:arcentroidave} shows the annual averages of these four centroids. Evident in the plots are the expected high latitudes at the beginning of each cycle and the steady migration of the emerging activity towards the equator during each cycle.  Curves similar to these were first shown (for 1952-1961) by Babcock (1961). Perhaps the most interesting new feature of Figure~\ref{fig:arcentroidave} is the alternating pattern of the positive and negative centroids in the two hemispheres. During Cycle~21 the negative centroids in the two hemispheres were on average displaced a few degrees north of their positive counterparts, then in Cycle~22 they were displaced a few degrees south. This pattern continued into the beginning of Cycle~23, when the negative centroids were again both displaced north of the positive centroids. These centroid displacements were statistically significant as the error bars show. Therefore throughout each cycle the active-region fields had a net poloidal component, pointing in the same direction in the two hemispheres. At the beginning of each cycle these poloidal fields opposed the poloidal field associated with the polar fields. During each cycle the polar fields reversed at the height of the activity maximum so that at the end of each cycle the active-region and polar-field poloidal components pointed in the same direction. This pattern is common to the two data sets. Because these centroid measurements were derived image by image, without comparing active region fields from different images, they are not seriously affected by the magnetograph-dependent differences between active region field measurements described above.

Around 2004 the pattern appears to have broken down. In the data from both observatories for both hemispheres the positive and negative centroids became very close to each other on average, and remained so until the beginning of the activity minimum around 2007. During 2007 and 2008 the negative centroids were closer to the equator than the positive centroids in both hemispheres, implying that the net active-region poloidal fields in the two hemispheres were opposed to each other. However, there was little active region flux during these two years. Since Cycle~24 began in earnest the centroids in the north hemisphere have reverted to $\pm 20^{\circ}$ latitudes but it is not yet apparent that significant centroid displacements have developed.
It is striking that the breakdown of the cyclical centroid displacements around 2004 coincided with the appearance of the asymmetry between the hemispheres in Figure~\ref{fig:butterfly}. It is not clear if there is a causal relationship between these two phenomena. The breakdown of the cyclical centroid displacements may also be associated with the weakness of the polar fields since they reversed in 2002-03. 

According to the Hale-Nicholson law (e.g., Priest~1982) the leading/following sunspots of bipolar active regions typically have opposite polarity in the two hemispheres and these polarities alternate from one cycle to the next. The Babcock-Leighton phenomenological model of photospheric flux transport relies on the active regions being preferentially tilted with respect to the east-west direction, such that on average the leading polarity is centered at a lower latitude than the following polarity according to Joy's law. The lead sunspots are also generally more intense and compact than the following sunspots. This net tilt of active regions provides the net poloidal field component that explains the cyclical reversal of the polar fields during each activity maximum according to the Babcock-Leighton model. Could the effective disappearance of net active-region poloidal flux be related to the failure of the polar fields to strengthen significantly after 2004 (Figure~\ref{fig:butterfly})? We will investigate this question in the next section where we compare the active region fields to the fields and field changes at high latitudes in Section~\ref{s:arstpl}.

\section{Comparison of active region fields to high-latitude streams and polar field changes}
\label{s:arstpl}

\begin{figure} 
\begin{center}
\resizebox{0.9\textwidth}{!}{\includegraphics*[10,230][600,530]{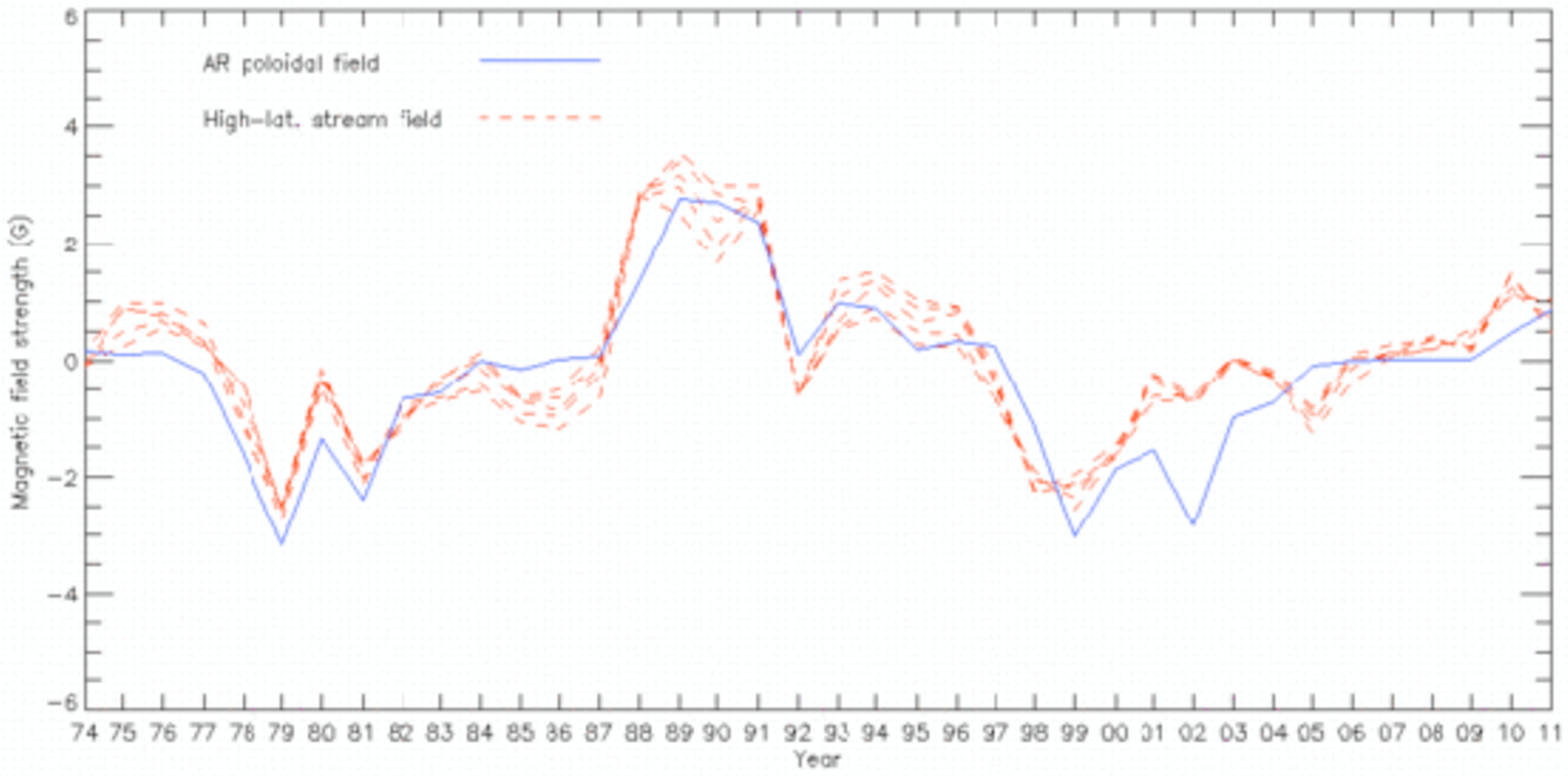}}
\resizebox{0.9\textwidth}{!}{\includegraphics*[10,230][600,530]{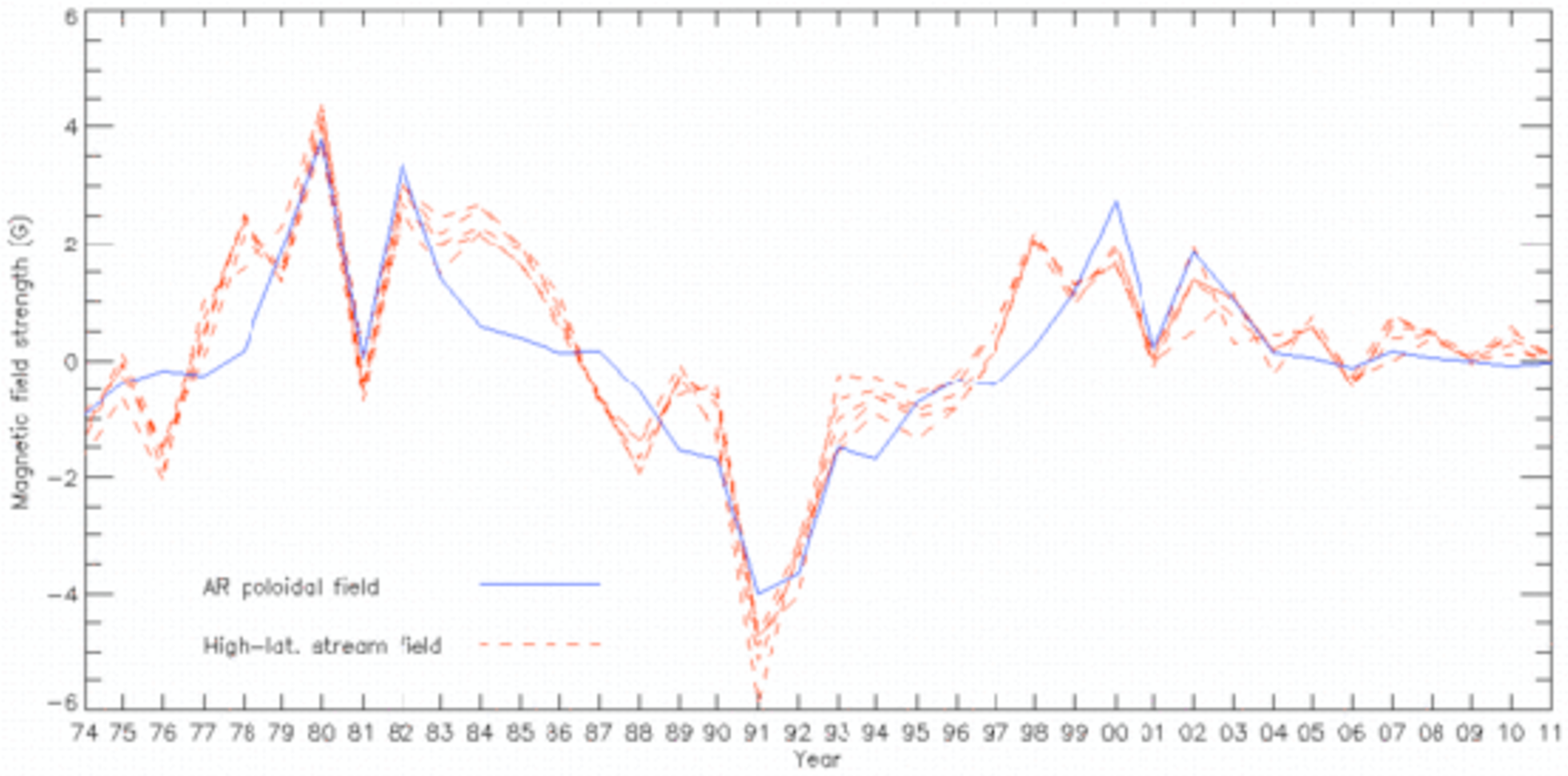}}
\end{center}
\caption{The proxy poloidal active-region flux (blue solid line) and the high-latitude stream field strengths at a selection of latitudes (red dashed lines) in the northern (top) and southern (bottom) hemispheres for the Kitt Peak data.}
\label{fig:arstreamave}
\end{figure}

\begin{figure} 
\begin{center}
\resizebox{0.9\textwidth}{!}{\includegraphics*[10,230][600,530]{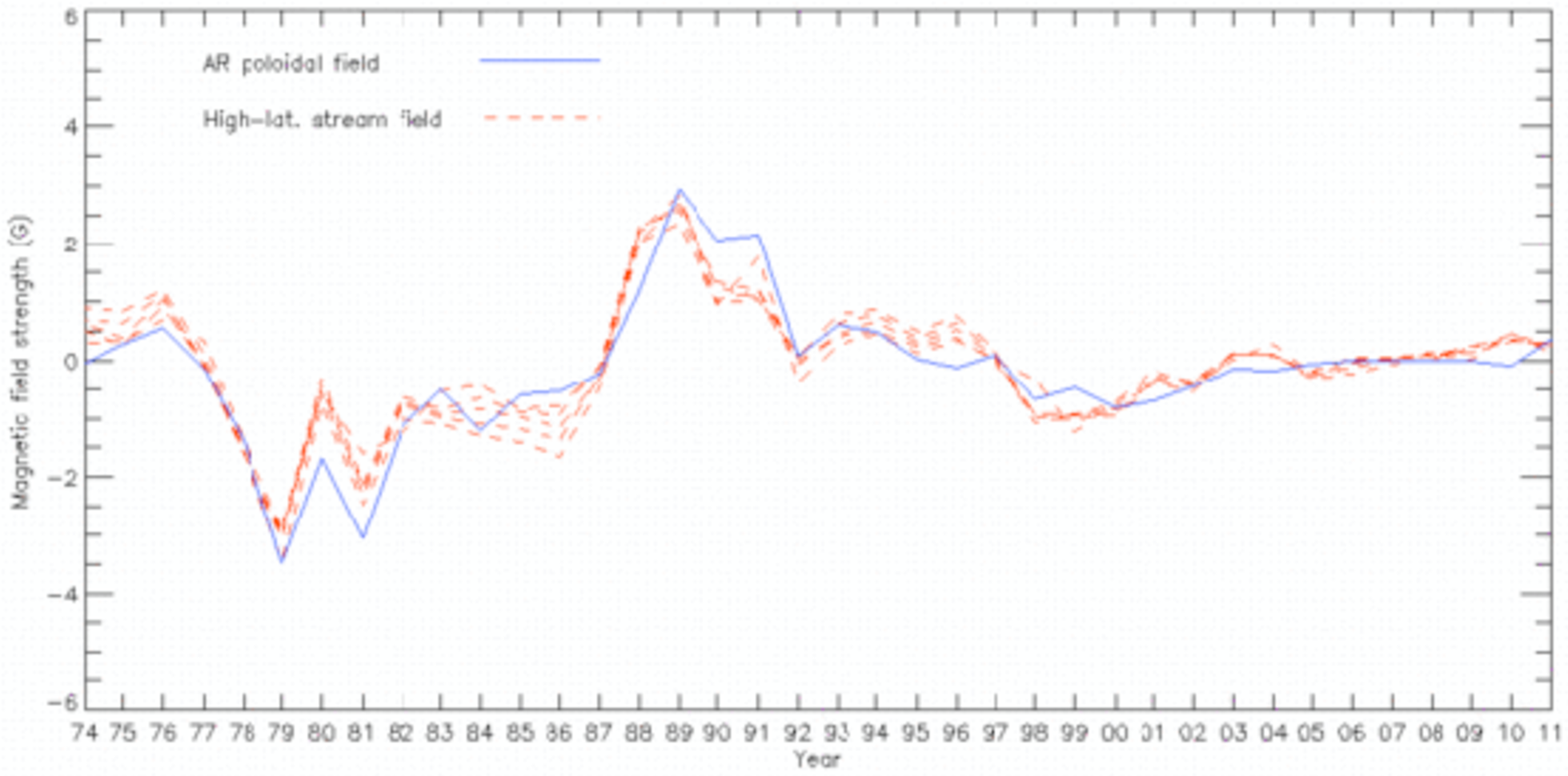}}
\resizebox{0.9\textwidth}{!}{\includegraphics*[10,230][600,530]{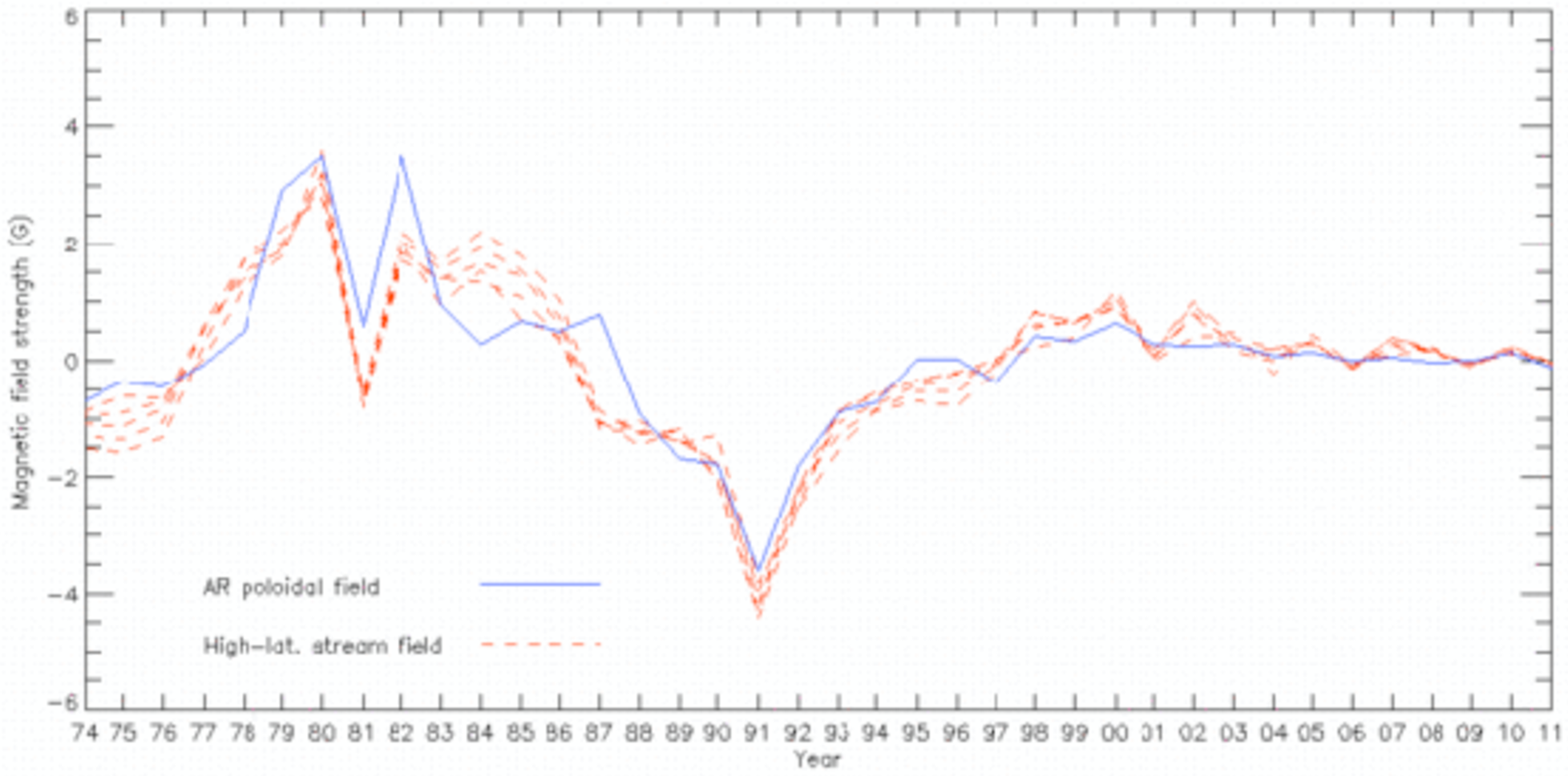}}
\end{center}
\caption{The proxy poloidal active-region flux (blue solid line) and the high-latitude stream field strengths at a selection of latitudes (red dashed lines) in the northern (top) and southern (bottom) hemispheres for the Mt. Wilson data.}
\label{fig:arstreamave_mwo}
\end{figure}

\begin{figure} 
\begin{center}
\resizebox{0.9\textwidth}{!}{\includegraphics*[10,230][600,530]{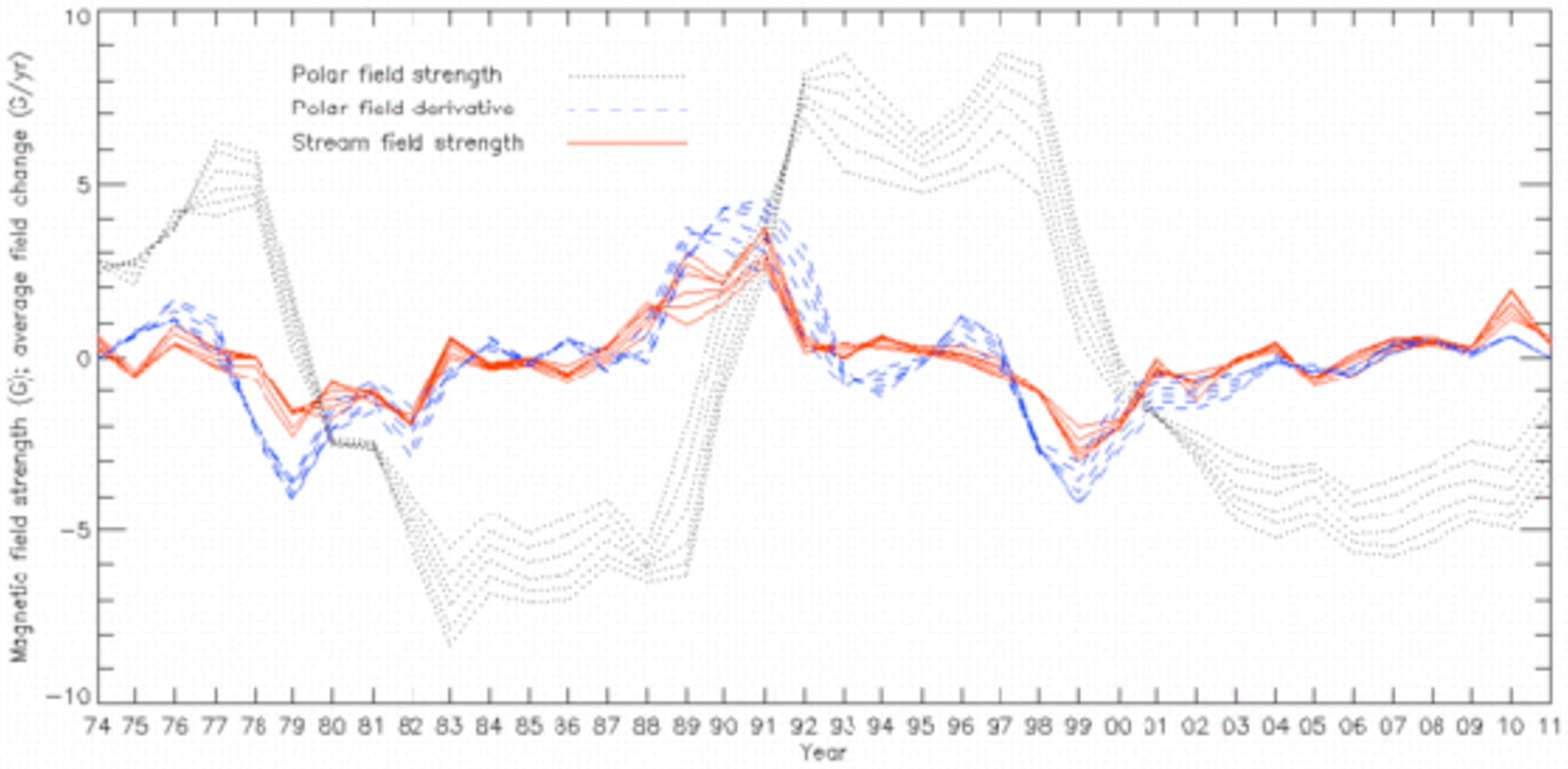}}
\resizebox{0.9\textwidth}{!}{\includegraphics*[10,230][600,530]{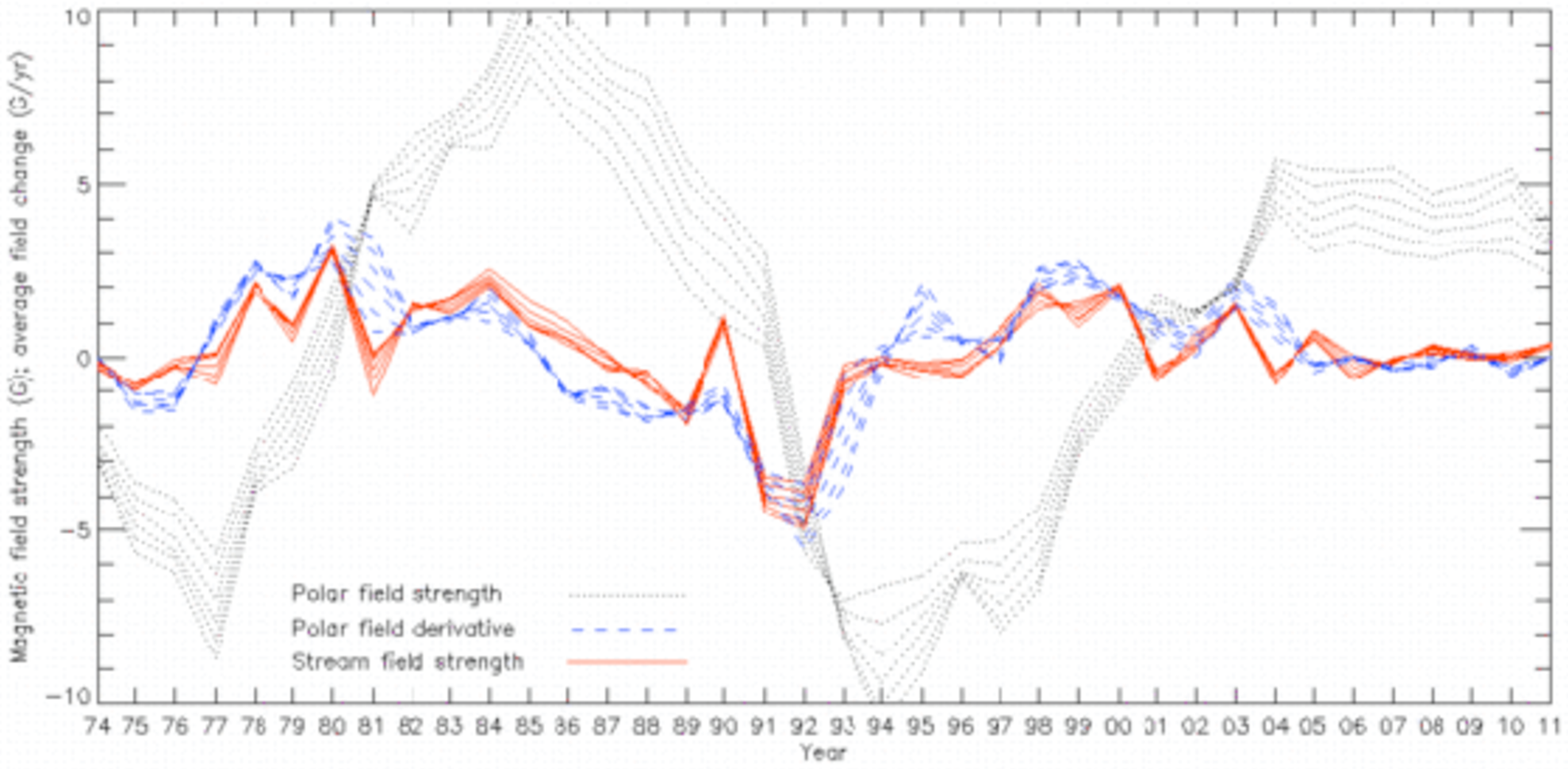}}
\end{center}
\caption{The polar field strengths (black dotted lines), their time derivatives (blue dashed lines) and the high-latitude stream field strength (red solid lines) at a selection of latitudes in the northern (top) and southern (bottom) hemispheres for the Kitt Peak data.}
\label{fig:polestreamave}
\end{figure}

\begin{figure} 
\begin{center}
\resizebox{0.9\textwidth}{!}{\includegraphics*[10,230][600,530]{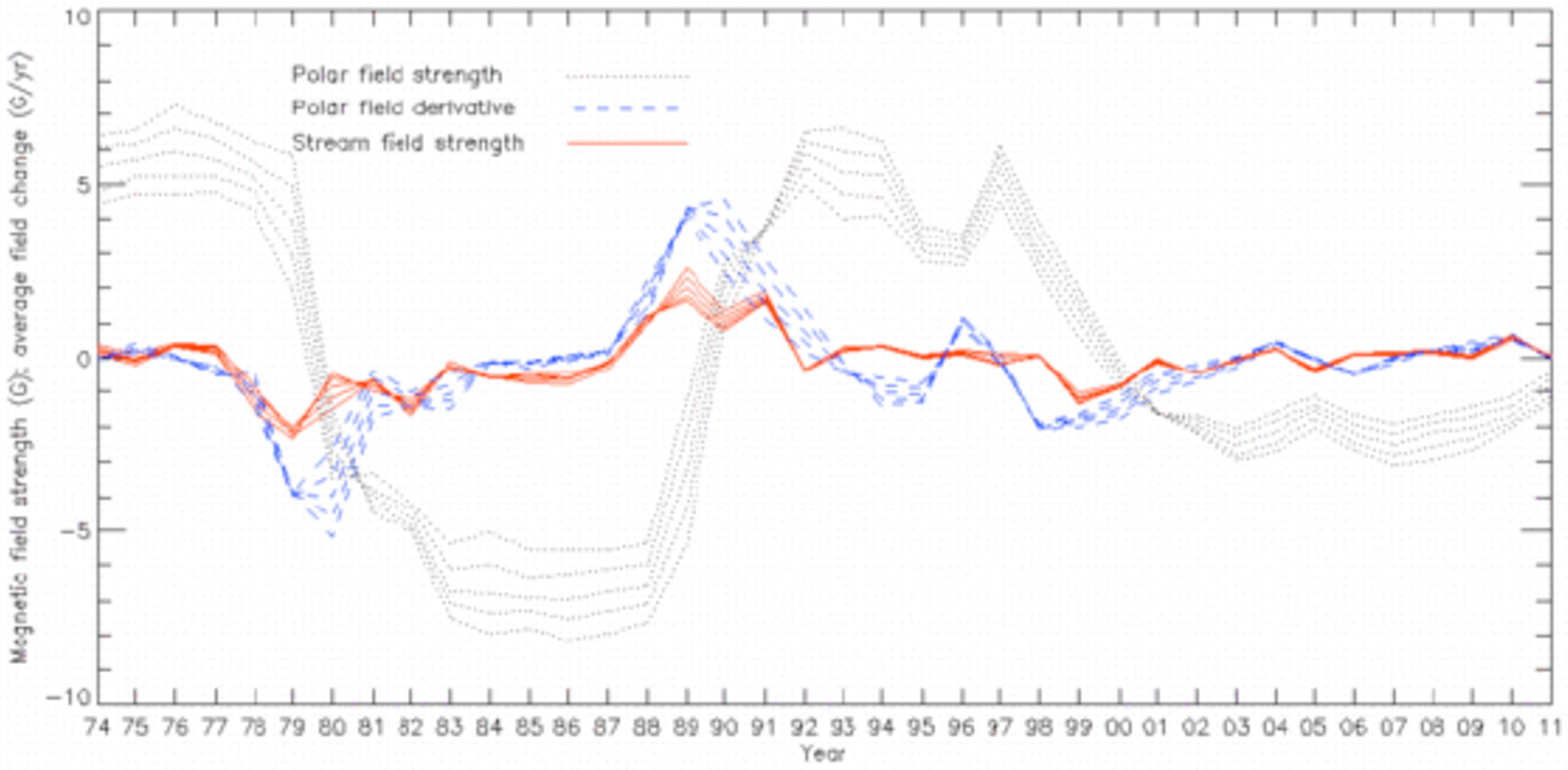}}
\resizebox{0.9\textwidth}{!}{\includegraphics*[10,230][600,530]{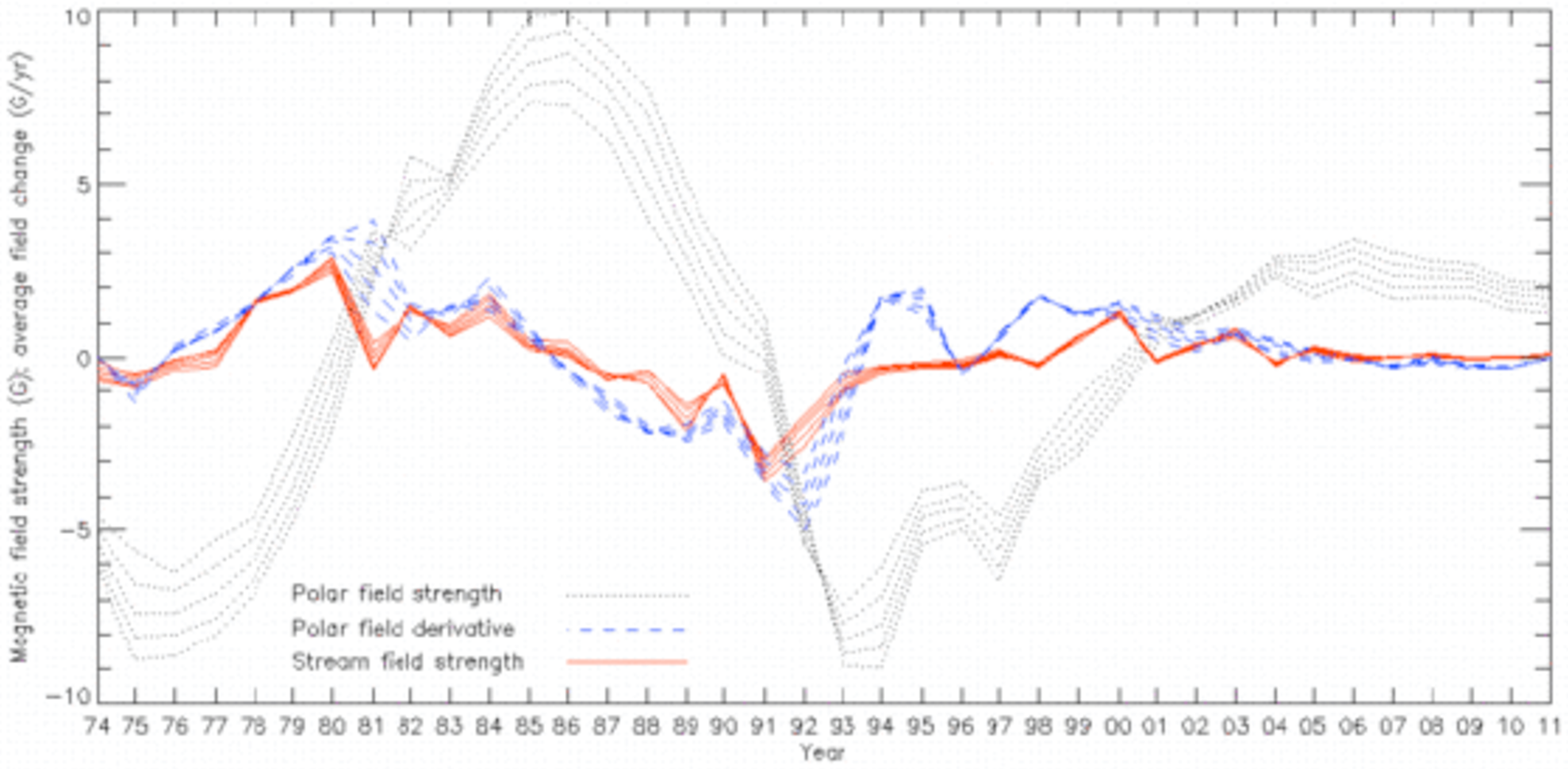}}
\end{center}
\caption{The polar field strengths (black dotted lines), their time derivatives (blue dashed lines) and the high-latitude stream field strength (red solid lines) at a selection of latitudes in the northern (top) and southern (bottom) hemispheres for the Mt. Wilson data.}
\label{fig:polestreamave_mwo}
\end{figure}

The poloidal active-region flux is not directly measured in the line-of-sight observations used here, but we can construct a proxy for this quantity. We expect the poloidal component of the active region flux in a given hemisphere to be proportional to the total active flux in the hemisphere and also to be proportional to the signed displacement between the positive and negative centroids in the hemisphere. We therefore represent the poloidal component of the active-region flux in each hemisphere by a proxy for this quantity, derived by multiplying for each hemisphere the total active region flux by the latitude distance between the positive and negative active region flux centroids. The active region poloidal field strengths are then estimated by dividing the fluxes by the estimated active area of the photosphere (between $0^{\circ}$ and $\pm 30^{\circ}$ in latitude). Not all of the net active region flux is believed to survive to reach high latitudes. Therefore we further divided the active region poloidal field strength by 3 for the Kitt Peak data and 2 for the Mt. Wilson data to produce curves of comparable size in Figures~\ref{fig:arstreamave} and \ref{fig:arstreamave_mwo}. The Kitt Peak results would imply that roughly a third of the active region net poloidal flux reaches high latitudes, and the Mt. Wilson results would imply that about a half does. The different results are likely to be due to the different spatial resolutions of the instruments at the two observatories. The active region proxy poloidal flux is plotted for each hemisphere in Figures~\ref{fig:arstreamave} (Kitt Peak data) and \ref{fig:arstreamave_mwo} (Mt. Wilson data). Also plotted are year-average field strengths at a selection of latitudes around $\pm 50^{\circ}$ corresponding to the poleward high-latitude flux stream patterns seen in Figure~\ref{fig:butterfly}. There is striking similarity between this set of curves and the curves for the proxy active-region poloidal components. The two classes of field are correlated with Pearson correlation coefficient around 0.85 (north) and 0.87 (south) for the Kitt Peak data and 0.91 (north) and 0.88 (south) for the Mt. Wilson data. The correlation for the Kitt Peak northern hemisphere data is slightly affected by the large amount of active region flux detected during Cycle 23, shown in the bottom panel of Figure~\ref{fig:arflux}, which arises because the SOLIS/VSM instrument is capable of recording significantly stronger field strengths than its predecessors were capable of measuring. The southern Cycle 23 active region poloidal field (Figure~\ref{fig:arcentroidave}) is also a little affected in this way. Comparing the high-latitude stream fields in Figures~\ref{fig:arstreamave} and \ref{fig:arstreamave_mwo} with the total active region flux profiles in the bottom panels of Figures~\ref{fig:arflux} and \ref{fig:arflux_mwo}, it is clear that the total active region flux does not by itself correlate as well with the high-latitude stream fields as the proxy poloidal active region fields do. The varying active region latitude centroids play a major role in the correlation. 

The active-region flux carried poleward by the meridional flows are expected to cancel or accumulate at the poles, changing the field strength there, as Figure~\ref{fig:butterfly} indicates. We therefore search for correlation between the high-latitude flux transport patterns and the changes in the polar fields. The fields at polar latitudes ranging from about $\pm 63^{\circ}$ to about $\pm 70^{\circ}$ are plotted in Figures~\ref{fig:polestreamave} (Kitt Peak data) and \ref{fig:polestreamave_mwo} (Mt. Wilson data) with their time derivatives. There is no obvious correlation between the high-latitude streams and the polar fields. However, the time-derivatives of the polar fields do track the stream fields. The polar fields increase or decrease according to whether they are being fed like- or opposite-polarity flux by the poleward streams. The high-latitude stream flux and the rate of change of the polar fields are correlated with linear Pearson correlation coefficient around 0.83 (north) and 0.80 (south) for the Kitt Peak data and 0.80 (north) and 0.82 (south) for the Mt. Wilson data. Flux-transport processes have been simulated to model the polar fields in synoptic magnetograms (Worden and Harvey~2000, Schrijver, DeRosa and Title~2002) but correlations between measurements of active-region flux, high-latitude stream flux and polar field changes have not been reported before to our knowledge.

According to Figures~\ref{fig:polestreamave} and \ref{fig:polestreamave_mwo} the weak polar fields since around 2004 have been accompanied by net high-latitude stream field strengths close to zero. There are signs in these figures that the polar fields, particularly the north polar fields, are currently in the process of weakening and are perhaps beginning to reverse. This development coincides with a significant stream of positive northern high-latitude flux seen in the red curves in the top panels of Figures~\ref{fig:polestreamave} and \ref{fig:polestreamave_mwo} and also visible in the upper right corners of both panels of Figure~\ref{fig:butterfly}. The northern active region poloidal flux, shown in the top panels of Figures~\ref{fig:arstreamave} and \ref{fig:arstreamave_mwo}, also shows some evidence of this change. The lower right corners of both panels of Figure~\ref{fig:butterfly} show more subtle traces of a more recent surge of negative flux towards the south pole.

\section{The Chromosphere and High Photosphere}
\label{s:chromosphere}

\begin{figure} 
\begin{center}
\resizebox{0.99\textwidth}{!}{\includegraphics*[20,255][600,530]{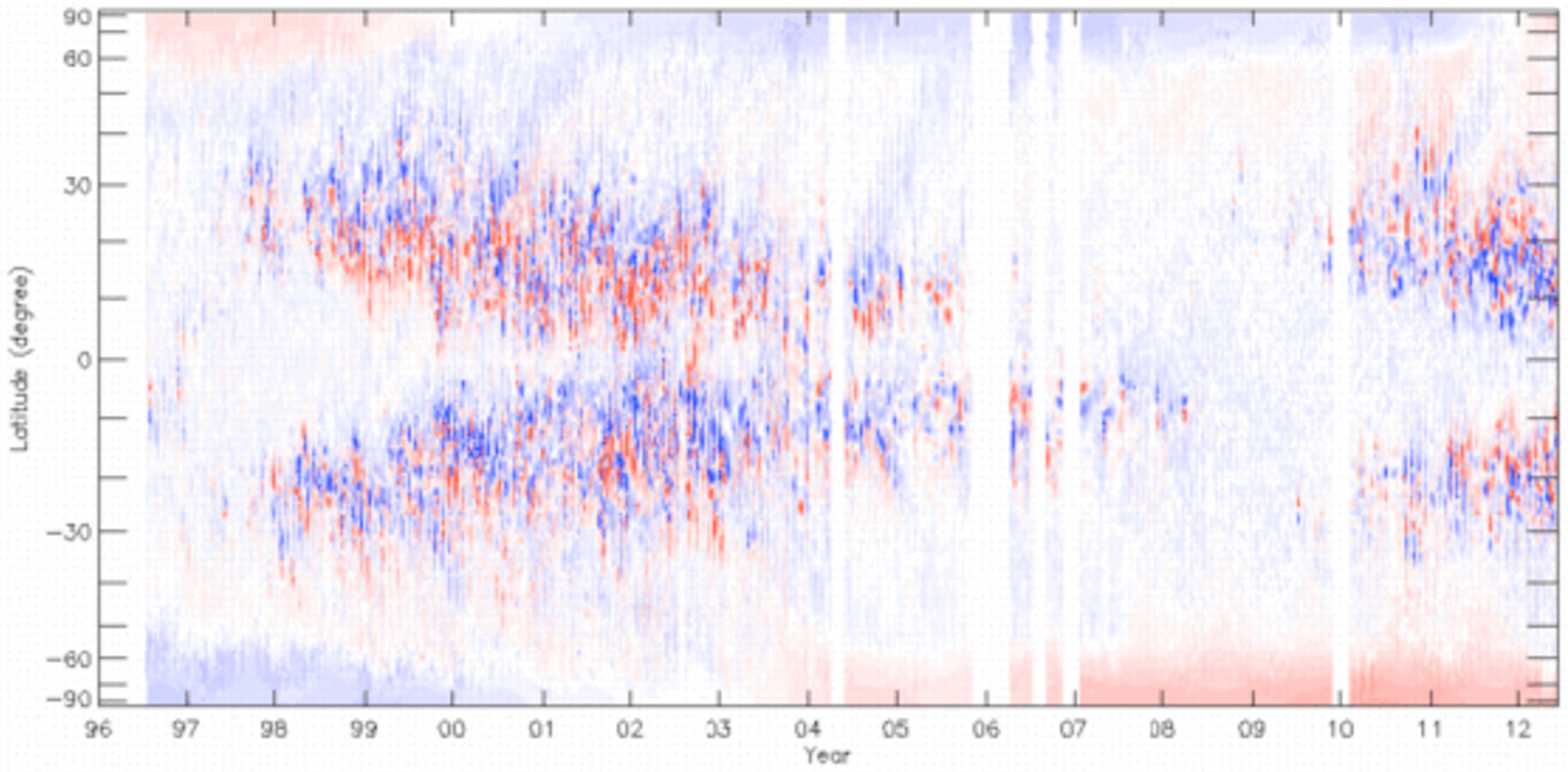}}
\resizebox{0.99\textwidth}{!}{\includegraphics*[20,255][600,530]{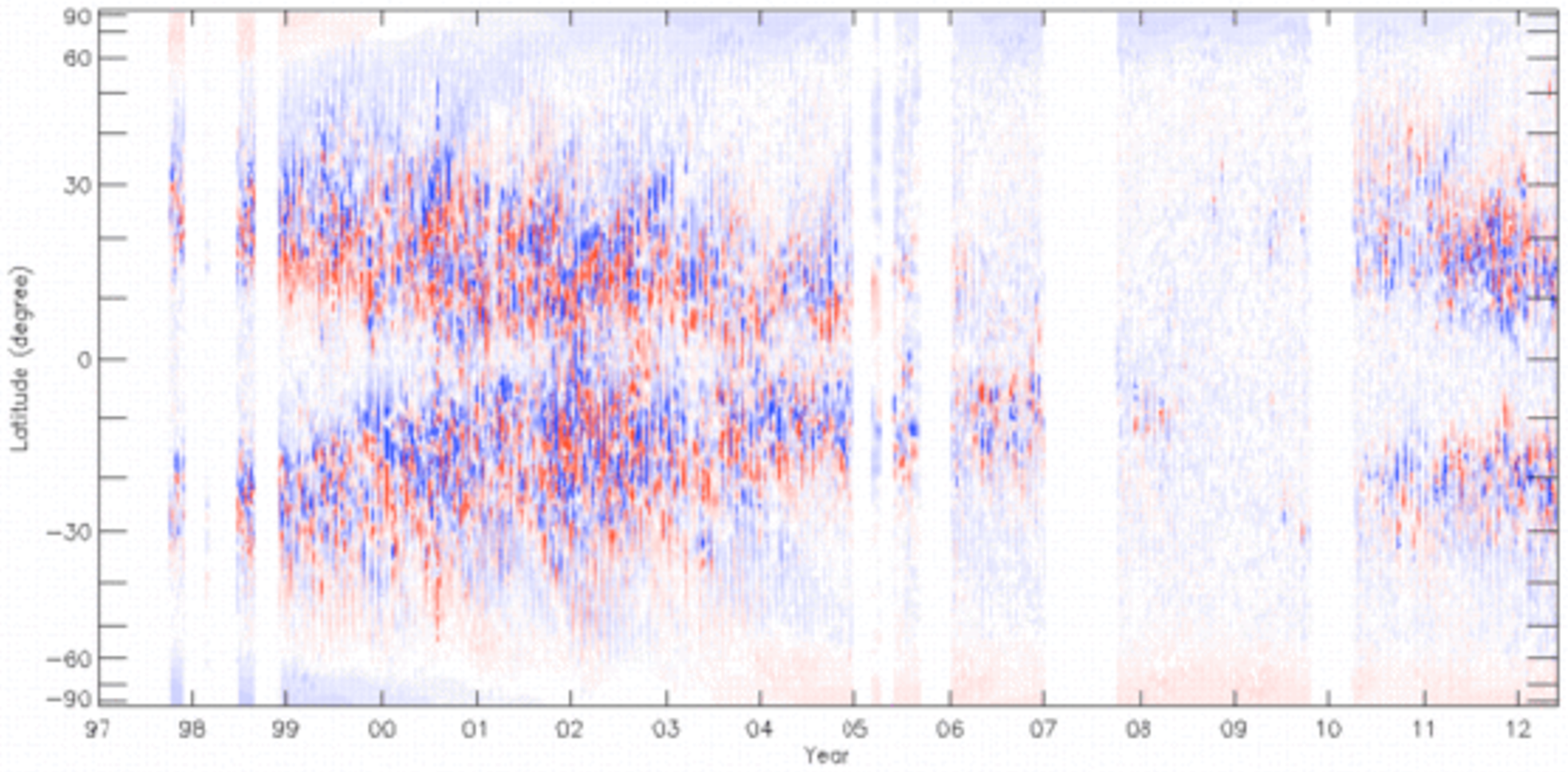}}
\end{center}
\caption{The butterfly diagram for the Kitt Peak (top) and the Mt. Wilson (bottom) chromospheric measurements. Each pixel is colored to represent the average field strength at each time and latitude. Red/blue represents positive/negative flux, with the color scale saturated at $\pm 20$~G.}
\label{fig:butterfly_chromo}
\end{figure}

\begin{figure} 
\begin{center}
\resizebox{0.9\textwidth}{!}{\includegraphics*[10,230][600,530]{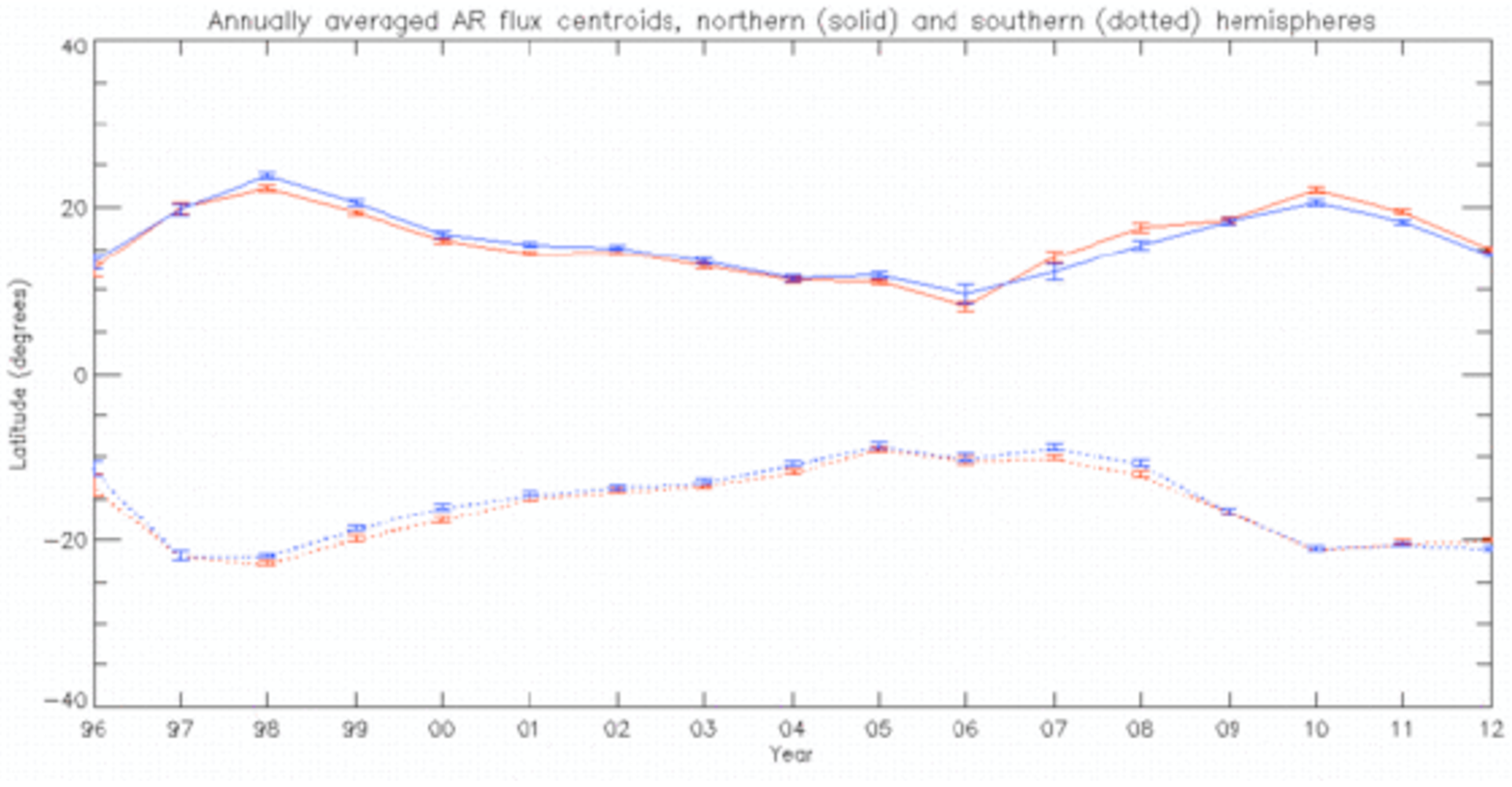}}
\resizebox{0.9\textwidth}{!}{\includegraphics*[10,230][600,530]{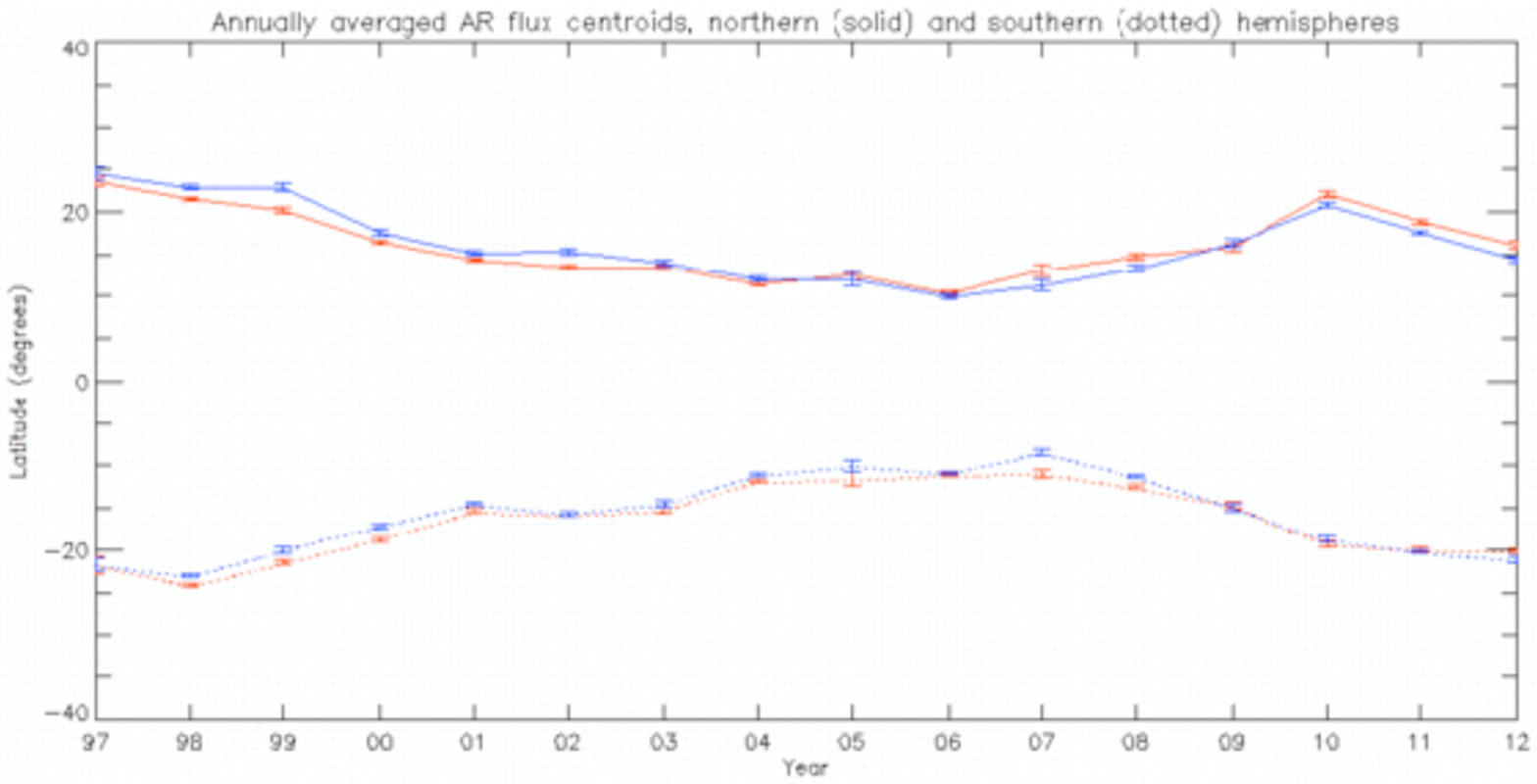}}
\end{center}
\caption{The latitudinal centroids of the active fields as functions of time for positive (red) and negative (blue) polarities in each hemisphere for the chromospheric measurements from Kitt Peak (top) and the Mt. Wilson (bottom).}
\label{fig:arcentroidave_chromo}
\end{figure}

As well as the photospheric observations, the SPMG and SOLIS instruments have been taking measurements in the Ca~{\sc II} 854.2~nm chromospheric spectral line since 1996. Meanwhile, the Mount Wilson 150-foot tower magnetograph has been simultaneously observing in the photospheric Fe~{\sc I} line at 525.0~nm and the  Na~{\sc I} D1 line at 589.6~nm since 1997. Although the latter is often regarded as a chromospheric line, most of the Stokes V signal is formed in the wing of the line profile and corresponds mostly to the high photosphere (Uitenbroek~2011). For the purposes of this paper, we will refer to the Na~{\sc I} D1 measurements from Mt. Wilson as ``high-photospheric'' and keep the term ``photospheric'' for the Fe~{\sc I} measurements described in the previous sections. We show butterfly diagrams of the chromospheric and high-photospheric measurements from the two observatories in Figure~\ref{fig:butterfly_chromo}. The high-latitude fields in Figure~\ref{fig:butterfly_chromo} are weaker than those in Figure~\ref{fig:butterfly} because a radial field correction was applied to the photospheric data but not to the chromospheric and high-photospheric data. We discussed in Section~\ref{s:butterfly} that the photospheric field is approximately radial. Magnetic fields measured using spectral lines formed higher in the atmosphere have been found to behave differently, with no preferred direction (Petrie \& Patrikeeva~2009). Unlike in photospheric measurements, in chromospheric full-disk images sunspots and network elements appearing unipolar near disk center acquire a ``false'' opposite polarity on their limb-ward side as they move limbward. This is a signature of a highly inclined, expanding field structure forming canopies at chromospheric heights (Jones~1985). Pietarila et al.~(2010) have interpreted the near-limb apparent bipolar magnetic features seen in high-resolution Hinode SOT NFI NaD1 observations using a simple flux tube model, and found evidence that magnetic features with vastly varying sizes have similar relative expansion rates.

The fact that the photospheric field is approximately radial allows us to approximate the full magnetic vector field strength there by dividing the observed longitudinal field component by the cosine of the heliocentric angle (the angle between the line of sight and the local radial vector). No equivalent approximation is available for fields higher in the solar atmosphere, making it difficult to estimate the strength of the full magnetic vector. The correlations between the different chromospheric and high photospheric fields are therefore poorer than for the photospheric fields. However the chromospheric data are useful because the polar longitudinal fields are stronger in the chromosphere than in the photosphere in general. This is because the polar fields expand super-radially, towards the observer, in the chromosphere. Hence we can see patterns in the chromospheric polar fields without applying a radial field correction. A stream of positive northern high-latitude flux can be seen in the upper right corners of both panels of Figure~\ref{fig:butterfly_chromo}, as well as more recent traces of a surge of negative flux towards the south pole. These correspond to patterns in Figure~\ref{fig:butterfly} described in Section~\ref{s:arstpl}.

Figure~\ref{fig:arcentroidave_chromo} shows the latitudinal centroids of the chromospheric fields as functions of time. In both hemispheres the positive and negative centroids are significantly closer together than the photospheric centroids shown in Figure~\ref{fig:arcentroidave}, before these began to converge around 2004. This may be due to the canopy structure in the chromosphere (and high photosphere). If the poleward/equatorward polarity of a bipole is tilted poleward/equatorward on average, then the centroid of the poleward/equatorward polarity will be displaced equatorward/poleward compared to the centroids of radial fields. This is because the longitudinal components of poleward/equatorward-tilted fields are stronger at lower/higher latitudes then the longitudinal components of radial fields.

\section{Conclusion}
\label{s:conclusion}

In 37 years of full-disk observations from the NSO Kitt Peak and Mt. Wilson 150-foot tower magnetographs covering several solar cycles, we have found statistically significant patterns consistent with the Babcock-Leighton phenomenological model for the solar cycle. 

The year-average centroids of the positive and negative active region fluxes showed statistically significant cyclical alternating patterns in each hemisphere, producing a net poloidal active region flux in both hemispheres of opposite sign, at the beginning of each cycle, to the poloidal flux of the polar fields. It was found that the annual averages of a proxy for the active region poloidal magnetic field strength, the magnetic field strength of the the high-latitude ($50-70^{\circ}$) poleward streams, and the time derivative of the polar field strength are all well correlated over both 37-year data sets in each hemisphere. The results also present two independent pieces of evidence that the active region net poloidal flux, having switched polarity from cycle to cycle in each hemisphere, in coordination with the polar fields according to the Babcock-Leighton model, effectively vanished in both hemispheres during Cycle 23. Since around the same time the polar fields have remained weak. As part of the larger pattern of correlation between the active region poloidal fields, high-latitude streams and polar field changes, this result links the weakness of the polar fields with the disappearance of the net poloidal active region field during Cycle 23. Jiang et al.~(2011) found that the weak polar fields of Cycle 23 could be reproduced in kinematic flux-transport dynamo models by a decrease in the average tilt angle of sunspots, but that this would also lead to a 1.5-year delay of the Cycle 23 polar field reversal that was not observed. Our results suggest that a decrease in the average tilt angle of bipolar active regions at the appropriate time during Cycle 23 ought to reproduce the Cycle 23 polar field reversal and weak field strength correctly (provided that the defining features of Cycle 23 are correctly represented in the model).

The weak polar fields of Cycle 23 have in the past been attributed to faster-than-average poleward meridional flows during Cycle 23 reported by, e.g., Ulrich (2010), Basu and Antia~(2010) and Hathaway \& Rightmire (2010). The basic argument is based on the Babcock-Leighton model. Fast meridional flows would inhibit trans-equatorial interaction between the leading polarities in the two hemispheres, essential for the Babcock-Leighton model to work, and thereby force the transported flux to be of such mixed polarity that the net effect of this flux on the poles would be small (e.g., Schrijver \& Liu~2008, Wang \& Sheeley~2009, Nandy et al.~2011). To significantly affect the polar fields, these fast meridional flows would have to occur at active latitudes, as Dikpati~(2011) has emphasized. Ulrich's~(2010), Basu and Antia's~(2010) and Hathaway and Rightmire's~(2010) meridional flow speed measurements do not support these arguments: they do not show evidence of significantly faster flows at active latitudes during Cycle 23 than during previous cycles.

The minimum of Cycle~23 was the longest and quietest in both Kitt Peak and Mt. Wilson data sets, but the active region fields have been becoming steadily stronger since 2009 with the northern hemisphere leading the southern hemisphere. The northern hemisphere is approaching maximum activity conditions in terms of maximum active region field strengths but is not so strong in terms of total active region flux. The southern hemisphere is not yet approaching maximum conditions at the time of writing. There was more hemispheric asymmetry in the activity level, as measured by total and maximum active region flux, during late Cycle 23 (after around 2004), when the southern hemisphere was more active, and during Cycle 24 up to the present, when the northern hemisphere has been more active, than at any other time since 1974. While it seems clear that the effective disappearance of the active region net proxy poloidal fields and the failure of the polar fields to become significantly stronger after 2004 are related, it is not clear if the hemispheric asymmetry, which began around the same time, was also connected. The cause of the vanishing active region poloidal fields during Cycle 23 remains to be investigated. Estimates by Schrijver \& Liu~(2008 and Stenflo \& Kosovichev~(2012) of tilt angles of large samples of Cycle 23 bipoles do not indicate a change in the average tilt angle during the cycle. On the other hand, Tlatov et al.~(2010) found distinctly different bipole tilt angle distributions for different classes of bipoles during Cycles 21-23, and Lef\`evre \& Clette~(2011) have found that the populations of sunspots of different sizes and degrees of complexity were distinctly different during Cycle~23 compared to previous cycles. To connect these results with the results reported here, the different contributions of distinct classes of active fields to the total active region poloidal flux need to be analyzed as they varied with time.

There are signs of high-latitude poleward streams of flux in the data, positive in the north and negative in the south, corresponding to weakening trends in the recent polar field measurements. These are indications that the polar fields are perhaps beginning to reverse. The speed, asymmetry and strength of this field reversal, and its relationship to the active region flux evolution, promise to teach us much about flux transport processes on the Sun.


%

%
\begin{acks}
I thank the referee for helpful comments. I thank Jack Harvey (NSO) and Roger Ulrich (UCLA) for help in accessing their data sets, for comments and corrections regarding the manuscript, and for article references. I thank Alexei Pevtsov and Giuliana de Toma for discussions and Luca Bertello for help in accessing the Mt. Wilson data set. SOLIS data used here are produced cooperatively by NSF/NSO and NASA/LWS. NSO/Kitt Peak 512-channel and SPMG data used here were produced cooperatively by NSF/NOAO, NASA/GSFC, and NOAA/SEL. This study includes data from the synoptic program at the 150-Foot Solar Tower of the Mt. Wilson Observatory. The Mt. Wilson 150-Foot Solar Tower is operated by UCLA, with funding from NASA, ONR and NSF, under agreement with the Mt. Wilson Institute.
\end{acks}

%

%
\bibliographystyle{spr-mp-sola-cnd} 


\IfFileExists{\jobname.bbl}{} {\typeout{}
\typeout{****************************************************}
\typeout{****************************************************}
\typeout{** Please run "bibtex \jobname" to obtain} \typeout{**
the bibliography and then re-run LaTeX} \typeout{** twice to fix
the references !}
\typeout{****************************************************}
\typeout{****************************************************}
\typeout{}}

\end{article} 
\end{document}